\documentclass[useAMS,usenatbib]{mnras}
\usepackage{cancel}
\usepackage{amssymb}
\usepackage{graphicx}
\usepackage{hyperref}
\usepackage{color}
\usepackage{BibDef}
\def\lsim{\mathrel{\rlap{\lower 3pt \hbox{$\sim$}} \raise 2.0pt \hbox{$<$}}}
\def\gsim{\mathrel{\rlap{\lower 3pt \hbox{$\sim$}} \raise 2.0pt \hbox{$>$}}}
\def\msun{\rm {M_\odot}}
\def\kms{\rm km\,s^{-1}}

\definecolor{Orange}{rgb}{1.0,0.5,0.15}

\title[Simplified galaxy formation with GIZMO] 
{Simplified  galaxy formation with mesh-less hydrodynamics}
\author[A. Lupi et al.]{Alessandro Lupi,$^{1}$\thanks{E-mail:
lupi@iap.fr} Marta Volonteri,$^1$ and Joseph Silk$^{1,2,3,4}$\\
$^1$Sorbonne Universit\`{e}s, UPMC Univ Paris 6 et CNRS, UMR 7095, Institut d'Astrophysique de Paris, 98 bis bd Arago, F-75014 Paris, France\\
$^2$AIM-Paris-Saclay, CEA/DSM/IRFU, CNRS, Univ Paris 7, F-91191, Gif-sur-Yvette, France\\
$^3$Department of Physics and Astronomy, The Johns Hopkins University, Baltimore, MD 21218, USA\\\
$^4$BIPAC, University of Oxford,1 Keble Road, Oxford OX1 3RH, UK}
\begin{document}

\date{Accepted May 24th, 2017}

\pagerange{\pageref{firstpage}--\pageref{lastpage}} \pubyear{2017}

\maketitle

\label{firstpage}

\begin{abstract}
Numerical simulations have become a necessary tool to describe the complex interactions among the different processes involved in galaxy formation and evolution, unfeasible via
an analytic approach. 
The last decade has seen a great effort by the scientific community in improving the sub-grid physics modelling and the numerical techniques used to make numerical simulations more predictive. Although the recently publicly available code \textsc{gizmo} has proven to be successful in reproducing galaxy properties when coupled with the model of the MUFASA simulations and the more sophisticated prescriptions of the Feedback In Realistic Environment (FIRE) set-up, it has not been tested yet using delayed cooling supernova feedback, which still represent a reasonable approach for large cosmological simulations, for which detailed sub-grid models are prohibitive. In order to limit the computational cost and to be able to resolve the disc structure in the galaxies we perform a suite of zoom-in cosmological simulations with rather low resolution centred around a sub-L* galaxy with a halo mass of $3\times 10^{11}\,\rm \msun$ at $z=0$, to investigate the ability of this simple model, coupled with the new hydrodynamic method of \textsc{gizmo}, to reproduce observed galaxy scaling relations (stellar to halo mass, stellar and baryonic Tully--fisher, stellar mass--metallicity and mass--size). We find that the results are in good agreement with the main scaling relations, except for the total stellar mass, larger than that predicted by the abundance matching technique, and the effective sizes for the most massive galaxies in the sample, which are too small. 
\end{abstract}

\section{Introduction}
According to the current cosmological scenario, galaxies form when baryons cool and fall into the potential wells  of the dark matter haloes. While the theory of structure formation is now in good agreement with the observational constraints by \citet{planck16}, the evolution of the baryonic component is still poorly understood, in particular when we consider processes like star formation (SF), feedback by stars and active galactic nuclei, and the formation of massive black holes and their subsequent coevolution with the galaxy host. The ever-improving details in observations provide a better understanding of the physical processes regulating galaxy evolution, which in turn more tightly constrain theoretical models.

Progress in numerical techniques and larger computing power has also played a crucial role in improving the predicting ability of simulations thanks to higher resolution and more detailed modelling of physical processes. However, because of the vast dynamic range involved, many processes are still below the resolved scales and must be modelled using ad hoc sub-grid prescriptions. 
On the one hand, cosmological simulations of large volumes cannot reach very high resolution (below $\sim 100$ pc) and lack the ability to resolve the giant molecular cloud scales where stars form and supernovae (SNe) explode. They are however necessary in order to investigate a statistically relevant sample of galaxies \citep{vogelsberger14,schaye15,dubois14}. On the other hand, small-scale simulations can better capture physical processes, but are unable to take into account environmental effects on the system \citep[e.g.,][]{creasey13,federrath15,geen16,gatto17}. 
To bypass these limits , two options are normally adopted, calibrate semi-empirical models based on observational data, or use small-scale simulation results to distill sub-grid models which can be implemented into larger scale models. To date, several people have tried to improve current models for galaxy formation focusing on the neglected physics processes. As a generic example, \citet{agertz13,aumer13,creasey13,keller14,roskar14,kimc15} investigated the stellar feedback, \citet{pakmor13} study the role of magnetic fields and \citet{pakmor16,salem16,simpson16} modelled cosmic rays to accelerate galactic outflows.

The last decade has seen a strong effort by the scientific community in the development of these models, both for hydrodynamical simulations and semi-analytic models. The results obtained have demonstrated our ability to broadly reproduce the main galaxy properties \citep[e.g.,][]{vogelsberger14,somerville15,dubois16}, but we are still far away from a full understanding of the key processes. 

At the same time, the techniques used to solve the basic  hydrodynamical  equations and calculate gravitational interactions have shown great improvements. Up to a few years ago, only two techniques were commonly used by the galaxy formation community , the Lagrangian smoothed particle hydrodynamics (SPH) technique, where the fluid was sampled via a set of discrete tracers \citep[particles;][]{gingold82,katz89}, and the Eulerian adaptive mesh refinement technique, where the volume was discretized with a Cartesian grid able to adapt the cell size depending on the fluid properties \citep{teyssier02,bryan14}. Both techniques have  different advantages and limitations \citep{agertz07} and several attempts have been made to refine them.
Recently, two completely new approaches have been proposed with the codes \textsc{arepo} \citep{springel10} and \textsc{gizmo} \citep{hopkins15}. In the former, the simulated volume is distributed among a discrete set of particles (cells) using a Voronoi tessellation; in the latter, the volume is partitioned among particles using a kernel function, resulting in a set of unstructured cells with smooth boundaries. However, although these approaches are supposed to capture the advantages of both Lagrangian and Eulerian techniques, as demonstrated by the results achieved in the standard tests, their limits are still unclear, and they need to be  thoroughly analysed.

In particular, \textsc{gizmo} has  proven to be successful with the refined scheme of the FIRE set-up \citep{hopkins14,hopkins17,elbadry16} and with the sub-grid models of the MUFASA simulation \citep{dave16,dave17}, but it has not been used with delayed cooling SN feedback in a cosmological context.
In this study, we assess the performance of the new hydro scheme implemented in the code, coupled with a standard model for SF and delayed cooling SN feedback \citep{stinson06,teyssier13} in reproducing the evolution of a population of sub-L$_\star$ galaxies down to $z\sim0$. Despite being simple compared to the detailed models of the FIRE setup, or to simulations including additional physics, such as  magnetic fields, radiative effects and cosmic rays, the prescriptions we implemented in the code are still a reasonable approximation for large cosmological volumes, where the resolution is not as high as that reached with state-of-the-art zoom-in simulations, and worthy of being tested.

The paper is organized as follows. In Section \ref{sec:setup}, we present the code and the numerical setup of our simulations. In Sections \ref{sec:results} and \ref{sec:redshift} we analyse the results for the high-resolution runs at $z=0.5$ and the redshift evolution, respectively. Section \ref{sec:lowres} presents a parameter study on the low-resolution runs for the SF density threshold and the SF efficiency. Finally, in Section \ref{sec:conclusions} we discuss the limitations of the study and draw our conclusions.

\section{Simulation setup}
\label{sec:setup}
In this study, we perform a zoom-in simulation centred around a halo with $M_{\rm vir}=3\times10^{11}\rm\, \msun/h$ at $z=0$ with a violent merger history, starting from $z=100$ down to $z=0.5$, using the mesh-less finite mass (MFM) hydrodynamics scheme available in \textsc{gizmo}. We also compare the default setup with a similar one with a halo of similar virial mass ($10^{11}\,\msun$) with a quiescent merger history.

\subsection{Numerical technique}
\textsc{gizmo} \citep{hopkins15}, developed  from \textsc{Gadget3}, itself derivative from \textsc{Gadget2} \citep{springel05}, implements a new method to solve hydrodynamic equations, aimed at capturing the advantages of the two most commonly used techniques so far, i.e. the Lagrangian nature of SPH codes, and the excellent shock-capturing properties of mesh-based codes, and therefore avoiding their intrinsic limitations.
The code uses a volume partition scheme to sample the volume, which is discretized among a set of tracer `particles' which correspond to unstructured cells. Unlike moving mesh codes \citep[e.g., \textsc{arepo}][]{springel10}, the effective volume associated with each cell is not defined via a Voronoi tessellation, but is computed in a kernel-weighted fashion. Hydrodynamic equations are then solved across the `effective' faces among the cells using a Godunov-like method, as in standard Eulerian mesh-based codes. 
Gravity is based on a Barnes--Hut tree, as in \textsc{Gadget3} and \textsc{Gadget2}.  Fully adaptive gravitational softening for the various particle types have been implemented in the code, but in this study we rely on the standard approach of fixed gravitational softening.

For this study, we have implemented in the code common sub-grid models for radiative cooling, SF and supernova (SN) feedback. 
Radiative cooling is computed by means of the standardized chemistry and cooling library \textsc{grackle} \citep{kim14,bryan14}, run in {\it equilibrium} mode. Cooling rates for both primordial species and metals are provided by look-up tables pre-computed with the photoionization code \textsc{cloudy} \citep{ferland13}, as a function of density and temperature. Metal cooling rates are provided assuming solar abundances and then linearly rescaled with metallicity, which is followed as a passive scalar. We also include a uniform ultraviolet background following the model by \citet{haardt12}, already included in the \textsc{cloudy} tables.
SF is implemented following a stochastic prescription aimed at reproducing the local Schmidt--Kennicutt law \citep{kennicutt98}, where the SF rate is defined as
\begin{equation}
\rho_{\rm SF} = \epsilon \frac{\rho_{\rm g}}{t_{\rm ff}},
\end{equation}
where $\epsilon$ is the SF efficiency parameter, $\rho_{\rm g}$ is the local gas density and $t_{\rm ff}=\sqrt{\frac{3\pi}{32 G \rho_{\rm g}}}$ is the free-fall time. We enable SF only when gas particles match three criteria: (i) $\rho_{\rm g}> \rho_{\rm SF}$, where $\rho_{\rm SF}$ is the SF density threshold, (ii) $T< 2\times 10^4$ K, where $T$ is the gas temperature and (iii) $\nabla\cdot \mathbf{v}<0$, where $\mathbf{v}$ is the gas proper velocity. When the criteria are matched, we stochastically spawn a new stellar particle with $1/3$ of the progenitor gas cell mass.
Because of the low-mass and spatial resolution, our stellar particles correspond to an entire stellar population, following a Chabrier initial mass function \citep[IMF;][]{chabrier03}. According to stellar evolution theory, we consider three different processes for stellar feedback: Type II SNe, Type Ia SNe and winds by stars in the range $1-8\rm\, \msun$.
\begin{itemize}
\item After $\sim 4$ Myr, the most massive stars start to explode as Type II SNe. For stars between $8\rm\, \msun$ and $40\rm\, \msun$, we release $E_{\rm SN}=10^{51} \rm erg/SN$ via thermal injection on to the nearest 32 neighbour particles of the stellar particle corresponding to the kernel sphere of the particle, together with mass and metals. Due to the limited resolution, the energy-conserving phase of the bubble expansion cannot be followed in the simulation and the additional energy released would be rapidly lost because of radiative cooling, resulting in an ineffective SN feedback. In order to avoid gas from rapidly getting rid of this additional energy, we implement a delayed cooling prescription, where gas cooling is inhibited for the time needed to reach the momentum conserving phase \citet{stinson06}. This model has been proven to better reproduce the Schmidt--Kennicutt relation and the outflow mass-loading factor in isolated galaxy simulations compared to other more physically motivated models, as stated by \citet{rosdahl17}. The only limitation is that it produces `unphysical' high temperatures in a region of the density--temperature diagram where cooling is expected to be effective. Following the resolution-dependent approach of \citet{dubois15}, we shut off cooling for $t_{\rm delay}\sim 10$ Myr. We assume that Type II SNe return all their mass but $1.4\rm\, \msun$ and we follow metal production (via Iron and Oxygen yields whose production rates are thought to be metallicity independent) using the tabulated results by \citet{woosley07} fitted by \citet{kim14}. The total metal mass can then be defined as
\begin{equation}
M_{\rm Z} = 2.09M_{\rm Oxygen} + 1.06M_{\rm Iron},
\end{equation}
where $M_{\rm Oxygen}$ and $M_{\rm Iron}$ are the Oxygen and Iron mass, respectively. {The metal mass injected every timestep is computed convolving the yield function with the IMF.}
For stars above $40\rm\, \msun$ a black hole (BH) forms via direct collapse, thus we do not release either mass or metals.
\item Type Ia SNe, instead, occur in evolved binary systems, when one of the stars has become a white dwarf and has accreted enough mass to exceed the Chandrasekhar mass limit. Type Ia SNe explode according to a distribution of delay times, as described by \citet{maoz12}, scaling as $t^{-1}$ between 100 Myr and 1 Gyr after a burst of SF. Type Ia SNe leave no remnants and release into the environment $1.4\rm\, \msun$, $M_{\rm Iron}=0.63\rm\, \msun$ and $ M_{\rm Oxygen}=0.14\rm\, \msun$. Since Type Ia SNe occur a long time after the burst of SF, these events are usually located far away from the progenitor molecular cloud, and they are not clustered as Type II SNe. In this case, we release $10^{51}\,\rm erg/SN$, but we do not shut off cooling, as described in \citet{stinson06}.  
\item Because of their low masses, stars between 1 and 8 $\rm \msun$ evolve on longer time-scales compared to their massive counterparts and do not explode as SNe. During their evolution, they release part of their mass as stellar winds. We model stellar winds by injecting only mass and metals on to the particles neighbouring the stellar particle. Assuming the initial--final mass relation for white dwarfs by \citet{kalirai08}, the mass-loss for these stars can be computed as $w_{m}=0.394 + 0.109m\rm\, \msun$, where $m$ is the mass of the progenitor star. We assume that low-mass stars do not produce new metals, hence stellar winds only carry the progenitor star metallicity. 
\end{itemize}
The number of SNe (Type II and Type Ia) and the mass losses for low mass stars per timestep are determined according to the stellar lifetimes by \citet{hurley00}.
By taking into account these processes, the stellar feedback in our simulations is able to return $42\%$ of the stellar particle mass in a Hubble time, prolonging SF even in cases of no fresh gas inflows \citep[e.g.,][]{leitner11,voit11}.

In order to guarantee that the Jeans length is always resolved, we include an artificial pressure term  defined as (in proper units)
\begin{equation}
P_{\rm support} = N_{\rm J}^2 G \rho_{\rm gas}^2 \Delta x^2/\gamma,
\label{eq:psup}
\end{equation}
where $N_{\rm J}=4$ is the number of elements we want to resolve, $\gamma$ is the gas adiabatic index, $G$ is the gravitational constant, $\rho_{\rm gas}$ is the local gas density and $\Delta x$ is the size of the resolution element in the simulation. In our runs, this is set to the maximum of the gravitational softening length $\epsilon_{\rm gas}$ and the average interparticle spacing $\tilde{h}= (4/3{\rm \pi} h/N_{\rm Ngb})^{1/3}$, with $h$ the kernel size and $N_{\rm Ngb}=32$ the desired number of neighbours (the standard value used for the cubic spline kernel). The artificial pressure term replaces the particle real pressure used by the Riemann solver only when the Jeans length is not properly resolved. The choice not to replace the particle internal energy but only the pressure guarantees that the gas temperature can be evolved self-consistently.

In this set of simulations, we do not include massive black holes and their feedback. Black hole feedback is expected to play a sub-dominant role in low-mass  galaxies such as those we simulate \citep{dubois13}, with stellar feedback playing the dominant role \citep{dubois15}.

\subsection{Initial conditions}
The initial conditions are similar to those of the AGORA collaboration, using the same cosmological box of 60 Mpc/h edge and the same initial noise seed, but with a slightly larger high-resolution region centred on the target halo, in order to include a larger number of galaxies.
The central galaxy was chosen not to have galaxies more massive than half its mass within $2\times R_{\rm vir}$\footnote{The virial quantities have been defined using a density contrast of 360 with respect to the background density.} at $z=0$. The high-resolution region is initially computed as the smallest Lagrangian box in the initial conditions encompassing the particles falling within a sphere of 500 kpc around the central halo at $z=0$. We then slightly increase the Lagrangian box to include a larger number of galaxies, checking that no galaxies more massive than the central one enter in the high resolution region. The high-resolution box is then $4.2\times 4.7\times 6.5$ Mpc/h$^3$ comoving at $z=100$.  We generate our initial conditions using \textsc{Music} \citep{hahn13}, where we set the minimum level of the cosmological grid to 7, to get a minimum mass resolution of $7.7\times 10^9\, \msun/h$ and the maximum refinement level of the initial grid to 10 and 11 for the low- and high- resolution simulations, respectively, reaching a mass resolution of $1.2\times 10^7\,\msun/h$ and $1.5\times 10^6\,\msun/h$, respectively. Baryons are generated only at the highest resolution level, with an initial mass of $2.5\times 10^6\,\msun/h$ in the low-resolution simulations and $3.2\times 10^5\,\msun/h$ in the high-resolution simulations.
We adopt the $\Lambda$ cold dark matter  cosmological model consistent the Wilkinson Microwave Anisotropy Probe 7/9 results, where $\Omega_{\rm m}=0.272,\,\Omega_\Lambda=0.728,\,\Omega_{\rm b}=0.0455,\,\sigma_8 = 0.807,\,n_{\rm s}=0.961$ and $H_0=70.2\, \rm\kms Mpc^{-1}$ and we assume negligible contribution from both radiation and curvature. We start our simulations from $z= 100$ and evolve them down to $z=0$ for the low-resolution case and $z=0.5$ for the high-resolution one. For the low-resolution runs we explore the parameter space for SF, varying SF density threshold and the SF efficiency to assess how galaxy properties are affected. For the high-resolution runs, instead, we vary the SF density threshold and the merger history of the halo,
the latter in order to check the particular choice for the initial conditions. The high-resolution box for the quiescent merger history run is centred on to a halo with $M_{\rm vir}=1\times 10^{11}\,\msun$ at $z\sim0$, using the setup of the AGORA collaboration flagship paper \citep{kim14}.
 The full simulation suite is reported in Table~\ref{tab:suite}, where the quiescent merger history case is H20.0q and all the others run correspond to the violent merger history case, with the `L' prefix for the low-resolution runs and the `H' prefix for the high-resolution ones.

\begin{table}
\centering
\scriptsize
\caption{Description of the simulation suite. We show the run name in the first column, the SF density threshold in the second one and the SF efficiency in the third one. Columns 4 and 5 are the comoving softening lengths for dark matter (DM) and gas, respectively, which are used down to $z=9$, while the last two columns are the physical softening lengths for DM and gas used at $z<9$.} 
\label{tab:suite}
\begin{tabular}{lrrrrrr}
\hline
Name & $\rho_{\rm SF}$  & $\varepsilon_{\rm SF}$ & $\epsilon_{\rm dm} $ &$\epsilon_{\rm gas} $ & $\epsilon^{z<9 }_{\rm dm} $ & $\epsilon^{z<9}_{\rm gas} $\\
&$(\rm H/cm^3)$ & & $(c\rm kpc)$ & $(c\rm kpc)$ & $(\rm kpc)$ & $(\rm kpc)$\\
\hline
L0.2 & 0.2 & 0.01 & 10.0 & 2.0 & 1.0 & 0.2\\
L1.0 & 1.0 & 0.01 &10.0 & 2.0 & 1.0 & 0.2\\
L5.0 & 5.0 & 0.01 &10.0 & 2.0 & 1.0 & 0.2\vspace{0.1cm}\\

L1.0low &1.0 & 0.005 & 10.0 & 2.0 & 1.0 & 0.2\vspace{0.1cm}\\

H5.0 & 5.0 & 0.01 & 6.44 & 1.0 & 0.64 & 0.1\\
H20.0 & 20.0 & 0.01 & 6.44 & 1.0 & 0.64 & 0.1\vspace{0.1cm}\\
H20.0q & 20.0 & 0.01 & 6.44 & 1.0 & 0.64 & 0.1\\
\hline
\end{tabular}
\end{table}

\section{Methods used to extract the information from our simulations and limits in the comparison with observations}
\label{sec:limits}
We describe here the methods we use to extract information from our simulated data and the validity and possible biases when comparing it with observations. To identify the galactic haloes in our simulations, we use the \textsc{amiga halo finder} (\textsc{ahf}) tool \citep{knollmann09}. 

For all the analyses reported here but the halo profile, we only consider the haloes with at least 100 particles, a fraction of high-resolution dark matter particles $f_{\rm high}>96\%$ and which formed stars during the simulations. We exclude all the sub-haloes identified by \textsc{ahf}, and include only central galaxies of main haloes. In order to avoid contamination from satellite galaxies in the comparison, stellar and gas properties are computed by considering the particles within $20\% r_{\rm vir}$ only\footnote{We checked that there was no appreciable difference in changing the aperture from 20\% to 50\% of the virial radius, and also in the case of a fixed aperture of 20~kpc.}.

\begin{itemize}
\item We measure the galaxy stellar masses by computing the total mass enclosed in a sphere of radius $20\% r_{\rm vir}$, while observations consider the total luminosity out to the maximum radius determined by the flux limit and then convert the luminosity into mass using a light to mass ratio. For instance, the stellar masses in \citet{beasley16}, which we use as a comparison with our simulations, are computed using the mass-to-light ratio from \citet{zibetti09}. We therefore expect our stellar masses to be biased towards slightly larger values.

\item For the galaxy sizes, we compute $R_{50}$ as the radius containing $50\%$ of the total stellar mass. Since the least massive galaxies in our sample show irregular structures, we bin the stellar mass in spherical shells. We will identify this method as `3D measure'. This is clearly different from what is done for observations, where a 2D surface brightness profile is fitted using a S\'{e}rsic model. In order to check the possible bias in our measure, we have decided to repeat the analysis by fitting the surface density profile of the galaxies with $M_\star> 4\times 10^9\,\msun$. Under the assumption of a constant mass-to-light ratio for the entire galaxy, this second approach has the advantage of being directly comparable with observations, although more prone to uncertainties for small/low-mass galaxies \citep{volonteri16,bottrell17}. 

\item Because of the relatively low resolution, not adequate to properly resolve the disc structure, except for the most massive galaxies in the sample, we estimate the rotational velocity $V_{\rm rot}$ of our galaxies as the circular velocity at $R=2R_{50,\rm\, bar}$, where $R_{50,\rm\, bar}$ is the radius encompassing $50\%$ of the total baryonic mass. The values computed with this method, as described by \cite{sales17}, show an approximately one-to-one correlation with the circular velocity measures at the radius enclosing $90\%$ of the H {\scriptsize I} mass in each galaxy, hence they are a good proxy to compare simulations with observations. In order to exclude the hot gas in the halo, we only consider gas below $10^4$ K as belonging to the galactic disc. 

\item To compute the stellar metallicity of the galaxies, we average the stellar particle metallicity in a mass weighted fashion. Observations, instead, use very different techniques, resulting in different systematics and biases compared to those of simulations. For instance, \citet{gallazzi05} use SDSS fibre spectroscopy, while \citet{kirby13} measures are based on single star spectroscopy and \cite{delgado14} uses an IFU survey, \textsc{CALIFA}. \citet{gallazzi05} showed that, although the SDSS fibre collected on average $\sim 30\%$ of the total light, the age and metallicity gradients varied weakly with the fibre aperture, suggesting that the metallicity estimates in the central region are a reasonable tracer of the total metallicity of the galaxy. As for \cite{delgado14}, the  measures are more consistent with the estimates from the simulations because the IFU technique maps the metallicity in the entire galaxy and also adopt mass weighting as we do. The only caveat is the different maximum radius used in simulations and observations. Finally, the approach of \citet{kirby13} is the most complex to compare, since we are unable to resolve single stars in the simulations, but only stellar populations. 

Moving to higher redshift, we compare the simulated data to \citet{maiolino08} and \citet{mannucci09}, who both use an IFU technique and \citet{yabe14}, who use a multifibre instrument on SUBARU, the same technique used by \citet{gallazzi05}. However, except for the different measure based on stellar mass instead of the spectral emission, the effect of the aperture size at higher redshift is less important because of the smaller galaxy sizes.
\end{itemize}

\section{The high-resolution runs: comparison with observations}
\label{sec:results}
In this section, we describe the results of our high-resolution simulation suite, comparing the properties of our simulated galaxies at $z\sim0.5$  with known scaling relations. 
We caution the reader that while a fairer comparison would require evolving the simulation down to $z=0.1$, we checked with our L-runs that the galaxy properties do not significantly change between $z=0.5$ and $z=0.1$. 

\subsection{Halo profiles}
\begin{figure}
\centering
\includegraphics[width=0.48\textwidth]{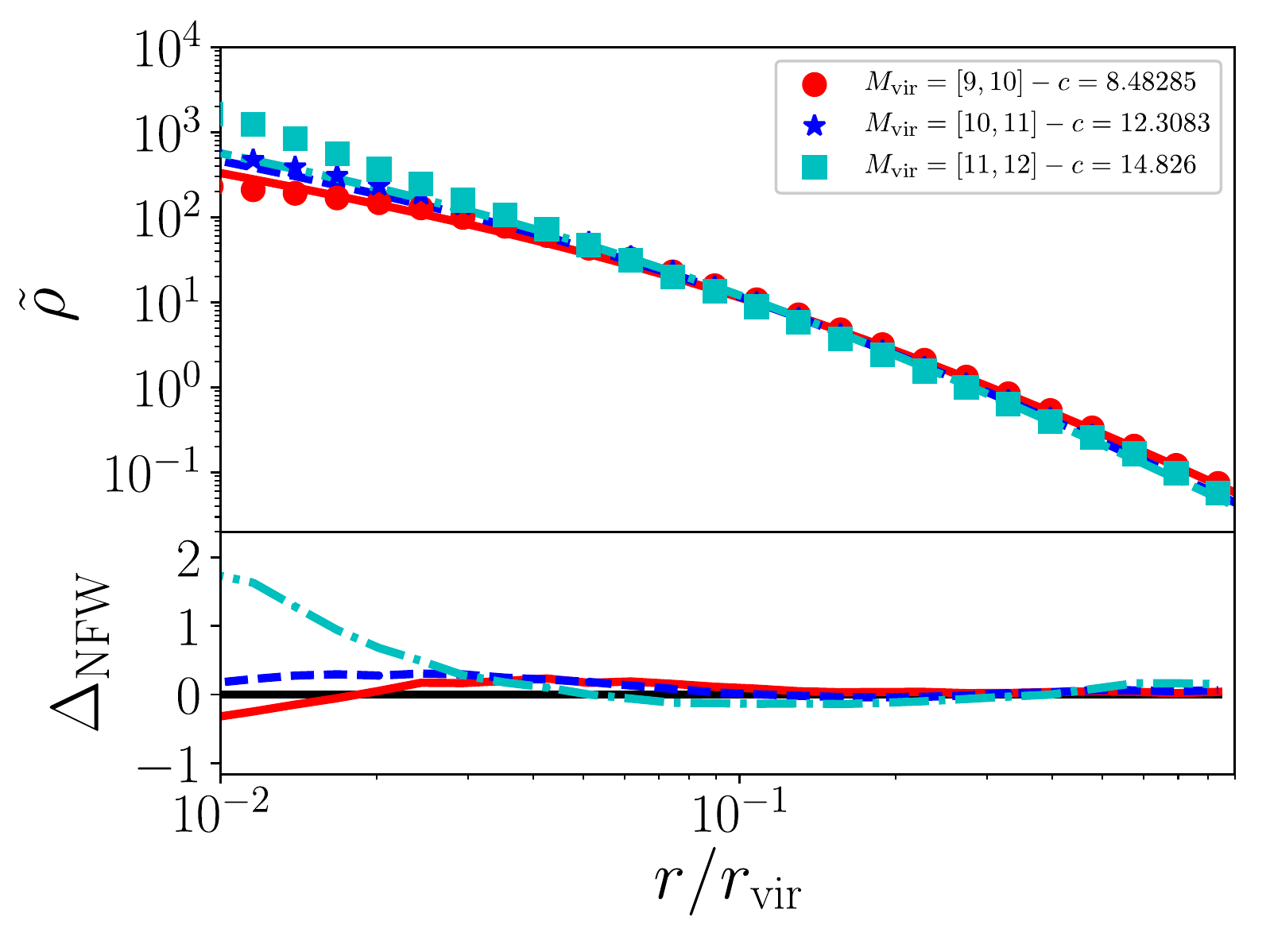}
\caption{Density profile of the haloes in our H-runs. The top panel shows the normalized density profile in three mass bins. Red dots, blue stars and cyan squares correspond to the measured profiles in the different mass bins, while the solid red, dashed blue and dot--dashed cyan lines are the best fit to the data of a NFW profile. In the bottom panel, instead, we show the deviation of the measured profile from the analytic one (in black), using the same line styles and colours of the top panel.}
\label{fig:haloprofile}
\end{figure}

For this analysis we consider all the haloes with at least 1000 particles, $f_{\rm high}>96\%$, and which are not identified as sub-haloes by \textsc{ahf}. We require a larger number of particles in the aim to better sample density profiles. We compute the halo density profile in spherical shells, in the range $0.01-1\times r_{\rm vir}$, where $r_{\rm vir}$ is the halo virial radius, and we normalize to unity both virial radii and total masses. In the top panel of Fig.~\ref{fig:haloprofile}, we show the halo median profiles in three mass bins between $10^9$ and $10^{12}\,\rm \msun$ and the corresponding fits to a Navarro--Frenk--White (NFW) profile. The number of haloes in the three bins is, respectively, $218, 69$ and $4$.
In the bottom panel, we plot $\Delta_{\rm NFW} = {\tilde{\rho} - \tilde{\rho}_{\rm NFW}}/{\tilde{\rho}_{\rm NFW}}$ for the three profiles. We can observe three different regimes, i.e. a central core for the least massive haloes, a well-behaved NFW profile for the intermediate range and an adiabatically contracted halo at the larger masses. At low masses, the central core can be produced via gravitational heating because of the SN driven fluctuations in the baryonic potential \citep{governato10,pontzen12,dicintio14}. At large masses, instead, the higher gas and stellar concentration in the centre is responsible for the adiabatic contraction of the halo, in agreement with recent studies by \citet{tollet16,peirani16}.

\subsection{Stellar to Halo mass relation}
\begin{figure}
\centering
\includegraphics[width=0.48\textwidth]{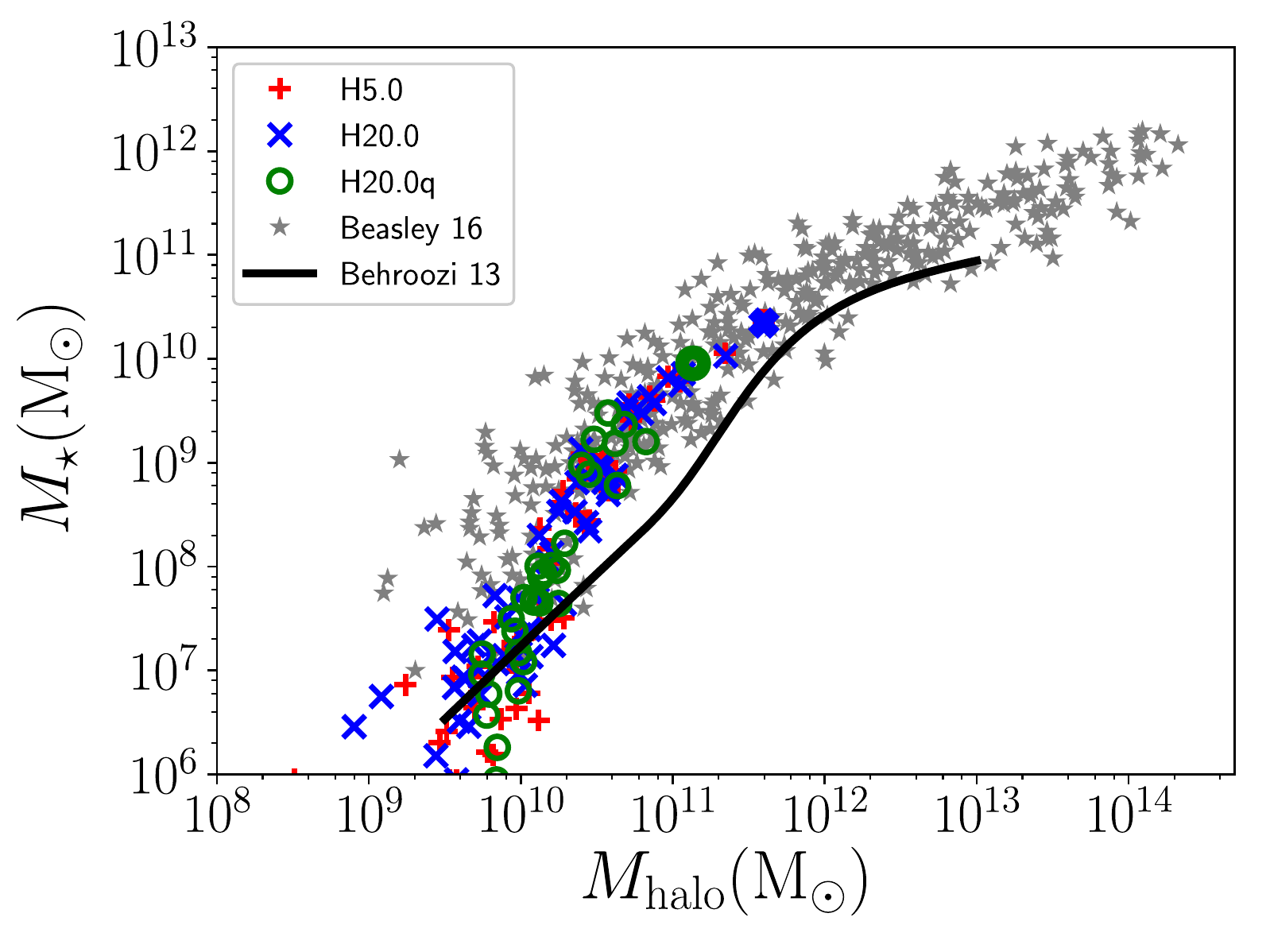}
\caption{Stellar to halo mass relation for our H-runs. The grey stars correspond to the observed data by  \citet{beasley16}, using the halo masses obtained via globular cluster counting, and the black line is the predicted relation from \citet{behroozi13c}. The most massive galaxies in our three simulations are the thicker markers in the plot.}
\label{fig:shmrelation}
\end{figure}

We now consider the relation between the galaxy stellar mass $M_\star$ and the halo mass, comparing our results with the model by \citet{behroozi13c} (B13 hereon) and with observations of Local Group dwarfs \citep{mcconnachie12} and nearby galaxies \citep{harris13}, using the halo mass estimates reported in \citet{beasley16}. Fig.~\ref{fig:shmrelation} shows the relation for the galaxies in our H-runs. 
The top panel shows the comparison between our simulations and nearby galaxy masses estimated via globular cluster counting \citep{beasley16}. We also plot the expected relation by B13 as a black solid line.

By comparing our data with \citet{beasley16}, we find very good agreement, with our galaxies lying well within the scatter. \citet{beasley16} also estimate halo masses by comparing their mass measurements to cumulative mass profiles of galaxies in hydrodynamical simulations. These masses are slightly higher than those obtained by globular cluster counting. Also in this case, our results are reasonably consistent with observational data, despite being shifted towards lower halo masses. The different density threshold used in the runs only moderately affects the total stellar mass, but does not have a significant impact on the general trend observed; its effect is more important for lower masses, dominating the scatter, while only a weak effect is visible for the most massive haloes. 

When we compare our data with the empirical model by \citet{behroozi13c}, instead, we get a reasonable match for the low-mass galaxies, while at high masses we overpredict the stellar mass by up to a factor of 10. This result suggests that the delayed cooling SN feedback, as implemented in the code, is able to reduce SF in the smaller mass haloes, but it is less effective at higher masses.

\subsection{Mass--size relation}

\begin{figure}
\centering
\includegraphics[width=0.48\textwidth]{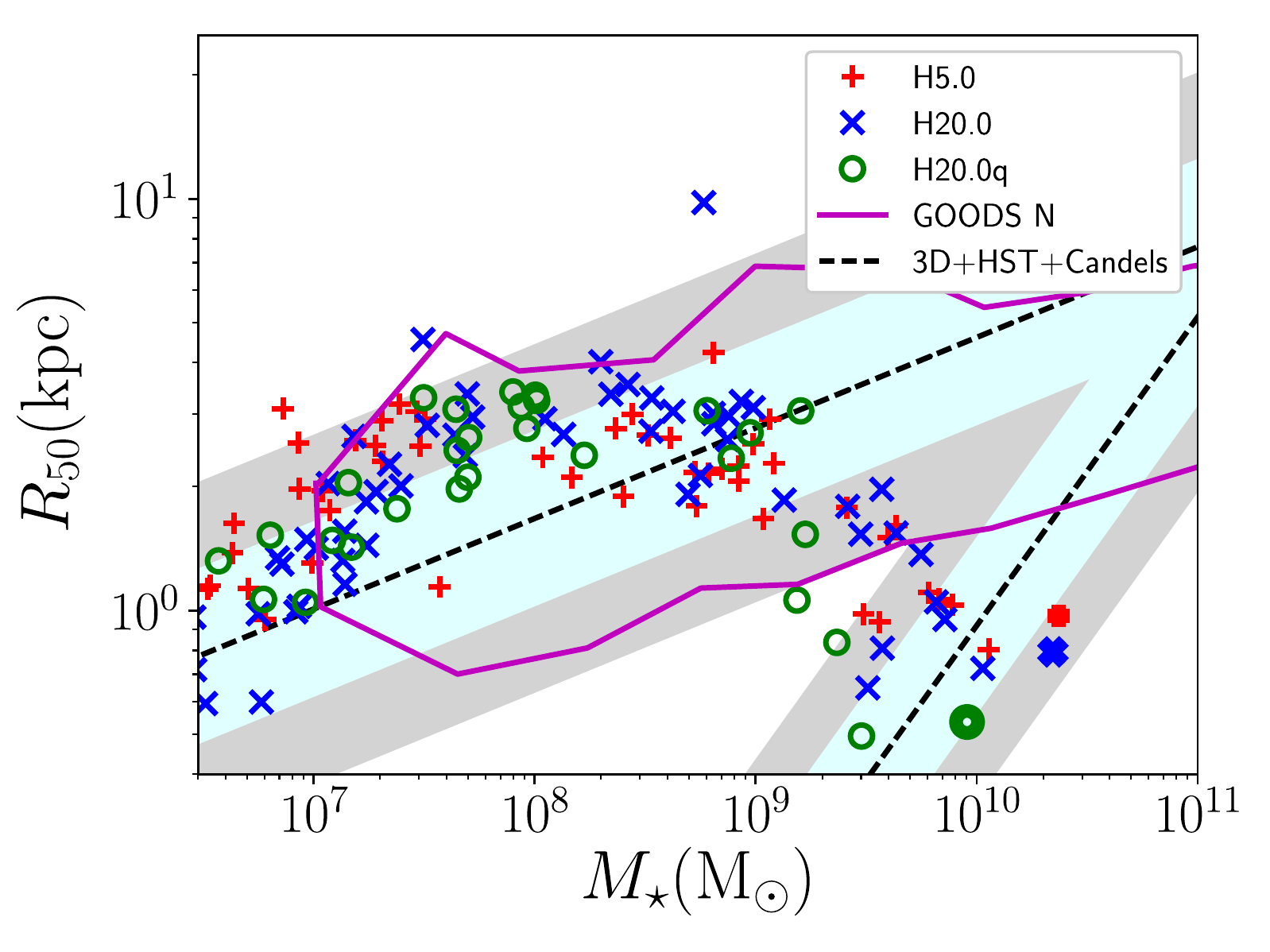}
\caption{Mass--size relation at $z=0.5$ for the our three high-resolution runs. The black lines correspond to the best fits from the 3D+HST+Candels survey \citep{vanderwel14} for both early (bottom curve) and late (top curve) type galaxies, with $1\sigma$ and $2\sigma$ uncertainties shaded in cyan and grey, respectively. The magenta contour, instead, corresponds to the data by \citet{ichikawa12} from the GOODS-N survey. The most massive galaxies in our three simulations are the thicker markers in the plot.}
\label{fig:msrelation}
\end{figure}

\begin{figure}
\centering
\includegraphics[width=0.52\textwidth]{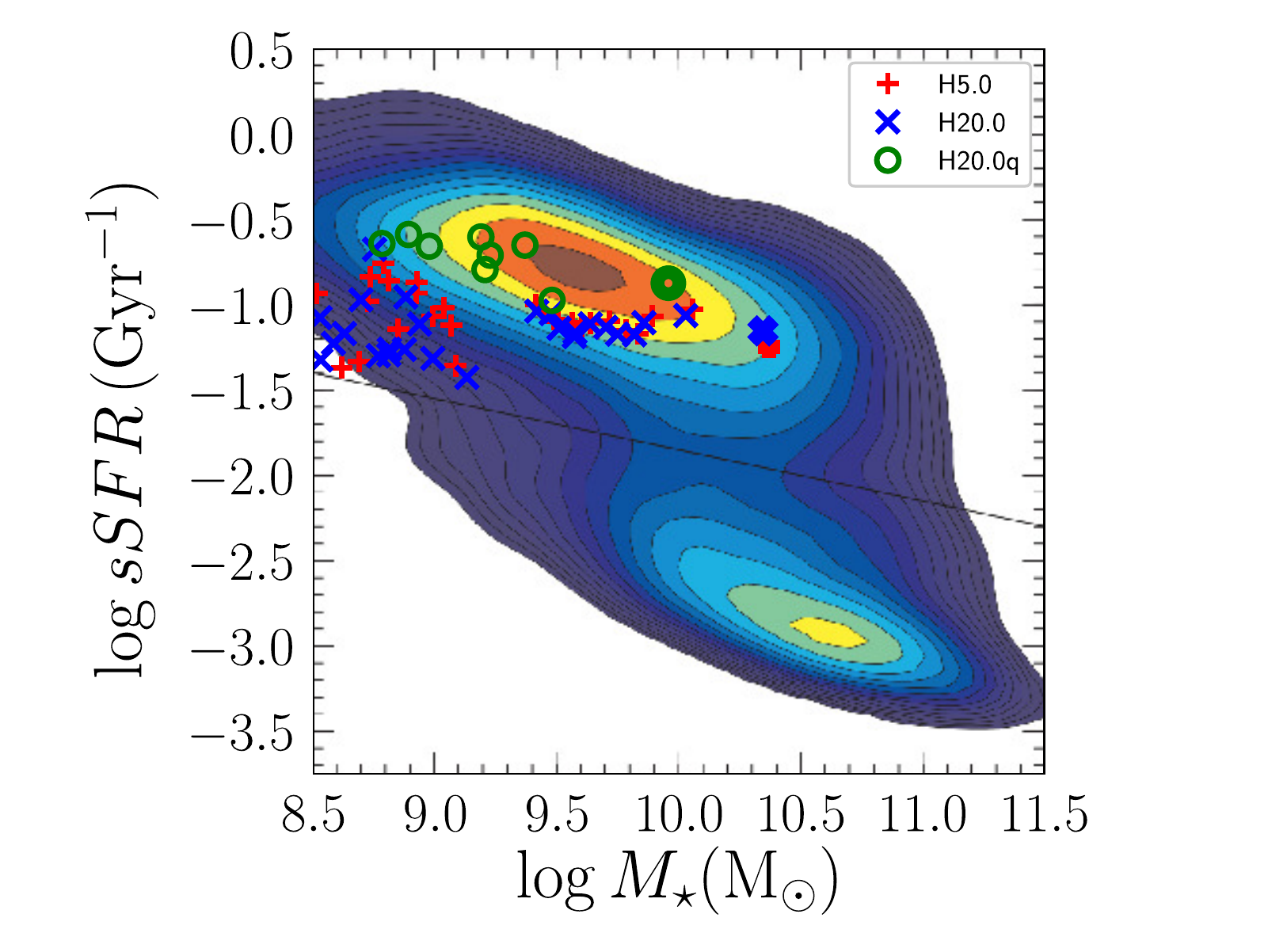}
\caption{sSFR versus stellar mass for our three high-resolution runs, compared with the data contours from \citet{knobel15}. The observational data correspond to a sample of ~123000 SDSS DR7 galaxies more massive than $10^9\,\msun$ (although the contour reported in their paper also show galaxies below $10^9\,\msun$) in the range $0.01 < z <0.06$.The values from the simulations are rescaled by $\left[(1+z_{\rm obs})/(1+z_{\rm sim})\right]^{-3}$ following \citet{lehnert15} to take into account the redshift difference because of the strong dependence of the sSFR with redshift. The most massive galaxies in our three simulations are the thicker markers in the plot.}
\label{fig:ssfr}
\end{figure}

Here we discuss the typical sizes of our galaxies obtained with the 3D measure, comparing them with observational data from the 3D+HST+Candels survey \citep{vanderwel14} and the GOODS-N survey \citep{ichikawa12} at $z\sim 0.5$. Fig.~\ref{fig:msrelation} shows the results of our H-runs, together with the best fit to the observational data in the 3D+HST+Candels survey (in black) and the contour from the GOODS-N survey (in magenta). We see a clear trend moving from low to high stellar masses. All the galaxies below $\sim 10^9\,\msun$ are fairly consistent within $1\sigma$ with observational data and show a direct correlation between size and stellar mass. However, the distribution shows a peak around $\sim 10^8\,\msun$ and then starts to bend towards smaller sizes. The most massive galaxies in our sample look rather compact compared to observations of late-type galaxies, but consistent with early-type ones.
In order to test whether these galaxies should really be considered early type, we estimated their specific SF rate (sSFR), which we define as ${\rm sSFR (1/yr)} = (M_{\star, <\Delta t}/\Delta t)/M_\star$, where $M_{\star, <\Delta t}$ is the mass of stars younger than $\Delta t= 50$ Myr, and $M_\star$ the total stellar mass of the galaxy, and compared it to the observational measurements by \citet{knobel15}. The observational data correspond to a sample of ~123000 SDSS DR7 galaxies more massive than $10^9\,\msun$ (although also the galaxies below $10^9\,\msun$ are shown in their plot) in the range $0.01 < z <0.06$. No other selection criteria have been applied to the data sample shown. The comparison with our simulated galaxies is reported in Fig.~\ref{fig:ssfr}, where the black line divides the SF galaxies (top part) from the quiescent ones (bottom part). Since the sSFR is expected to increase steeply with redshift up to $z\sim 2$, i.e. sSFR $\propto (1+z)^3$ \citep{lehnert15}, we rescale the sSFR in the simulated galaxies by a factor $\left[(1+z_{\rm obs})/(1+z_{\rm sim})\right]^3$, with $z_{\rm obs}\sim 0.03$ the average redshift of the observational data and $z_{\rm sim}=0.5$ the redshift of our simulations, assuming the stellar mass does not change significantly. According to the sSFR obtained, all the galaxies in the simulations should be of late type and significantly far from the dividing line for quiescent galaxies. The conclusion, which would have been the same even without applying the redshift scaling, is that the small sizes observed clearly deviate from expectations. 
If we compare the results of H20.0 with H5.0 in Fig.~\ref{fig:msrelation}, we observe that a higher density threshold produces slightly more massive central bulges in the most massive galaxies, skewing the relation towards smaller effective sizes. A separation between H5.0 and the runs with a higher density threshold can also be observed in Fig.~\ref{fig:ssfr}, although it is much smaller. This probably reflects the fact that the higher density threshold in H20.0 delays SF until gas has collapsed to higher densities, but as soon as the SF threshold is hit, the SFR becomes larger, resulting in a comparable mass and a slightly smaller size. In order to test this idea we compared the depletion time-scales for the five most massive galaxies in H20.0 and H5.0, binning the gas and the stellar distribution in cylindrical shells. We found that, while the gas density profile and the total stellar masses are very similar between the two runs, the depletion time-scales are very different. In the central kpc the values are comparable, with a ratio oscillating around unity, but at larger radii the ratio grows up to $\sim 4$, suggesting that the SF in H20.0 is suppressed at larger radii compared to H5.0 and confirming our idea.

\begin{figure}
\centering
\includegraphics[width=0.49\textwidth]{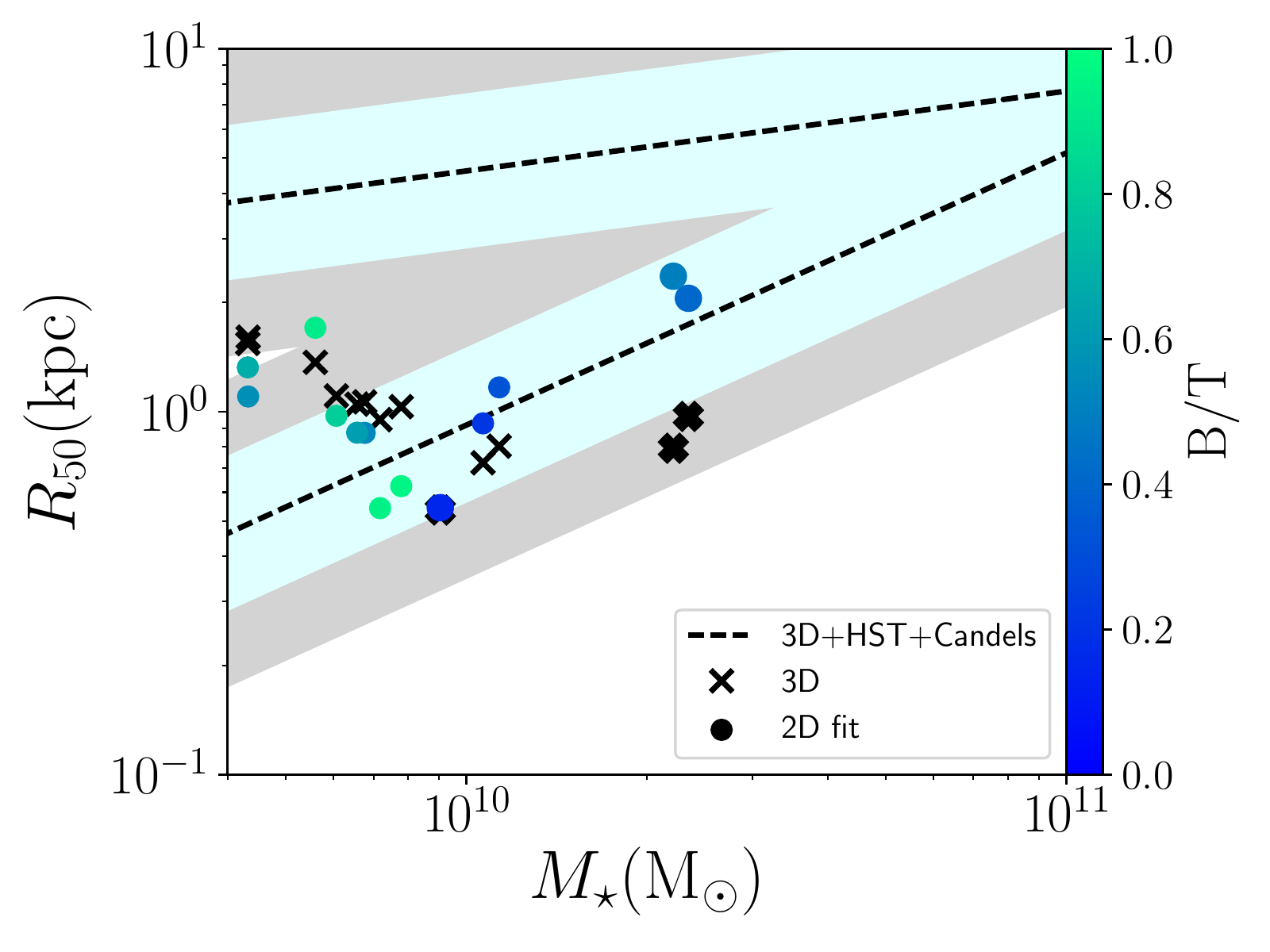}
\caption{Mass--size relation for galaxies with stellar masses larger than $4\times 10^9\,\msun$ in our H-runs, computed using the 3D measure (black crosses) and the 2D fit to the surface density profile using a S\'{e}rsic model. The circles correspond to the values obtained with the fit to the full profile and the stars to those obtained without considering the central kpc. The colours identify the bulge to total ratio (obtained by fitting the profile with two S\'{e}rsic models, for the bulge and the disc, respectively).}
\label{fig:b2tot}
\end{figure}

As discussed in \S\ref{sec:limits}, we have also tested the bias of our 3D measure by fitting the surface density profile of a sub-sample of our galaxies with a S\'{e}rsic model. The results of this analysis are reported in Fig.~\ref{fig:b2tot}, where we compare the data points from Fig.~\ref{fig:msrelation} with the sizes obtained with the 2D fit. We plot on the background the observational data by \citet{vanderwel14}. 
The 3D measures are shown as black crosses, while the 2D fit as blue-to-green dots, where the different colours correspond to different bulge to total (B/T) ratios, computed to assess whether our systems should be bulge or disc dominated\footnote{We have computed the B/T ratio by fitting the 2D surface density profile with two S\'{e}rsic models, one for the bulge, with index free to vary between 1 and 4 included and one for the disc with a fixed index of 1, respectively.}.
We do not see any clear trend in the comparison, with most of the galaxies scattered around the original data points, except for the four most massive galaxies considered, where we get an increase in size of up to a factor of 3. This difference moves the data points closer to the late-type relation (within $2\sigma$), but still more consistent with the regime of early-type galaxies.
Looking at the B/T ratio, we see that the most massive galaxies in the sample are disc dominated, and the bulge mass contributes at most for half the total stellar mass. This is opposite to what we observe for the less massive galaxies, which have B/T $> $0.5, but this is consistent with the fact that these galaxies show more irregular shapes and the contribution of the first component of the fit is dominant.

\begin{figure}
\centering
\includegraphics[width=0.48\textwidth]{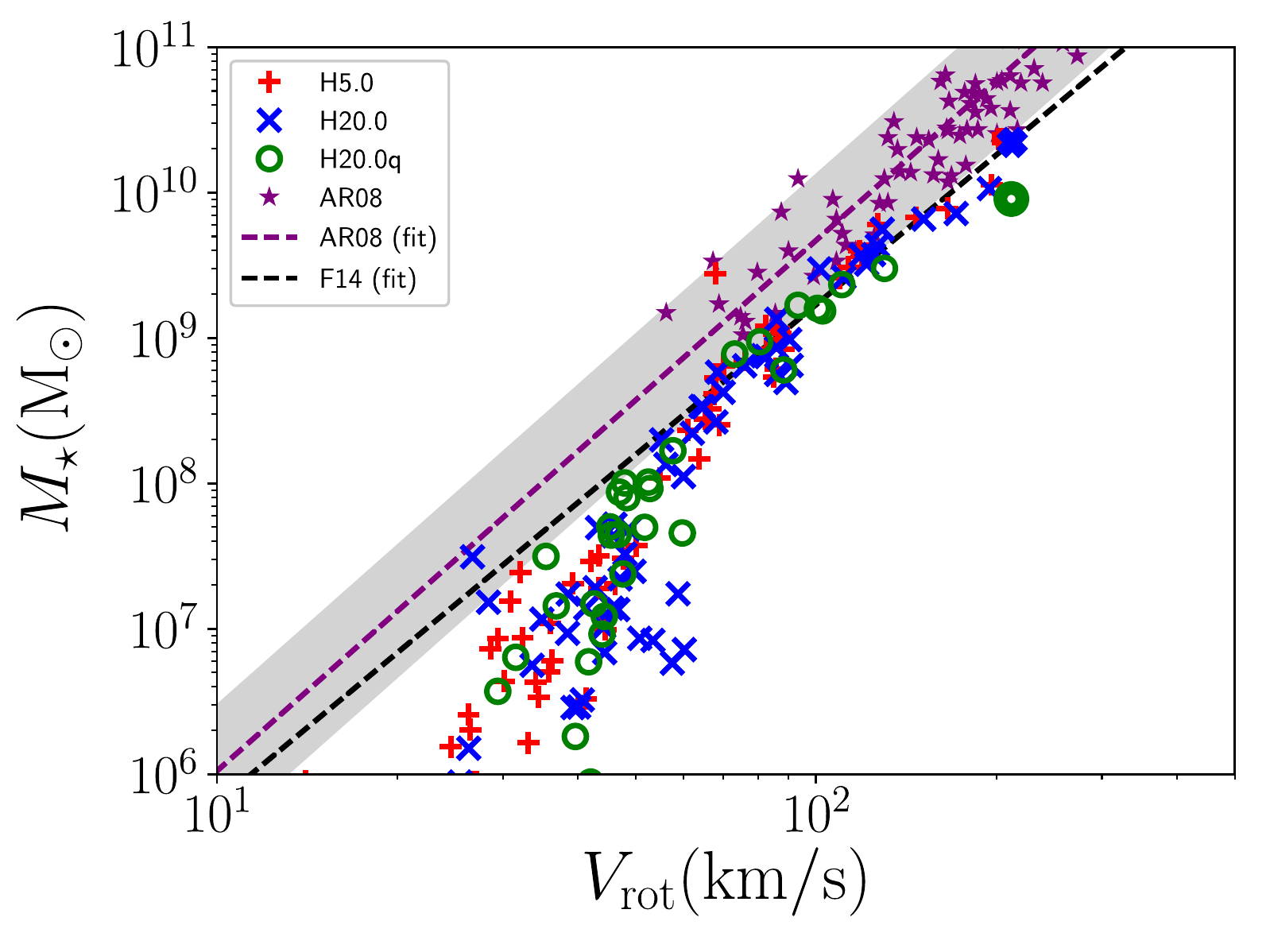}\\
\includegraphics[width=0.48\textwidth]{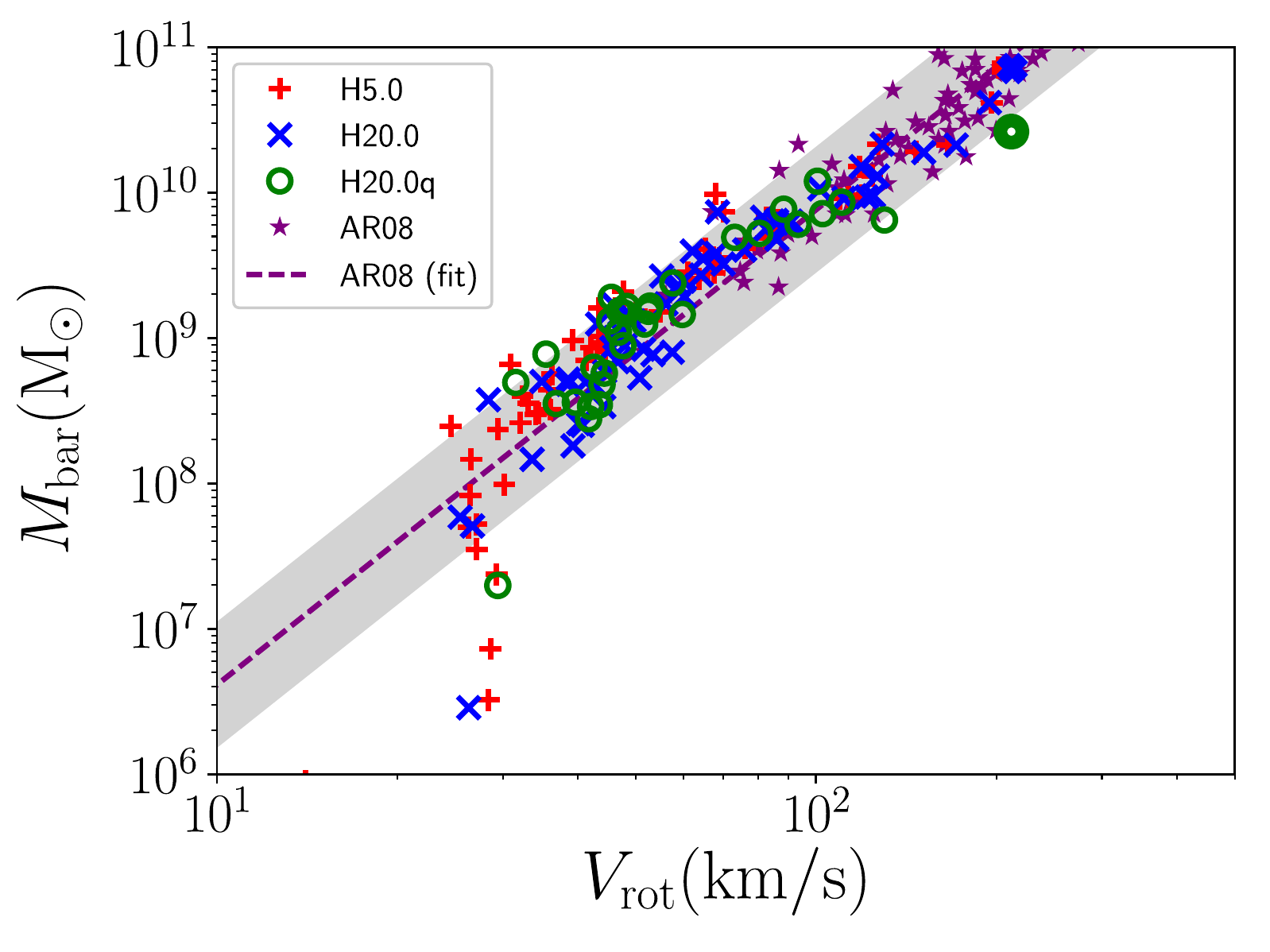}
\caption{Stellar (top panel) and baryonic (bottom panel) TF relations for our H-runs. The purple line corresponds to the orthogonal fit by AR08, with the $1\sigma$ scatter in grey, obtained from the data plotted as purple stars, and the black line in the top panel is the fit by F14 for galaxies at $z \in [0.45, 1.0]$. The most massive galaxies in our three simulations are the thicker markers in the plot.}
\label{fig:tf}
\end{figure}

\begin{figure}
\centering
\includegraphics[width=0.48\textwidth]{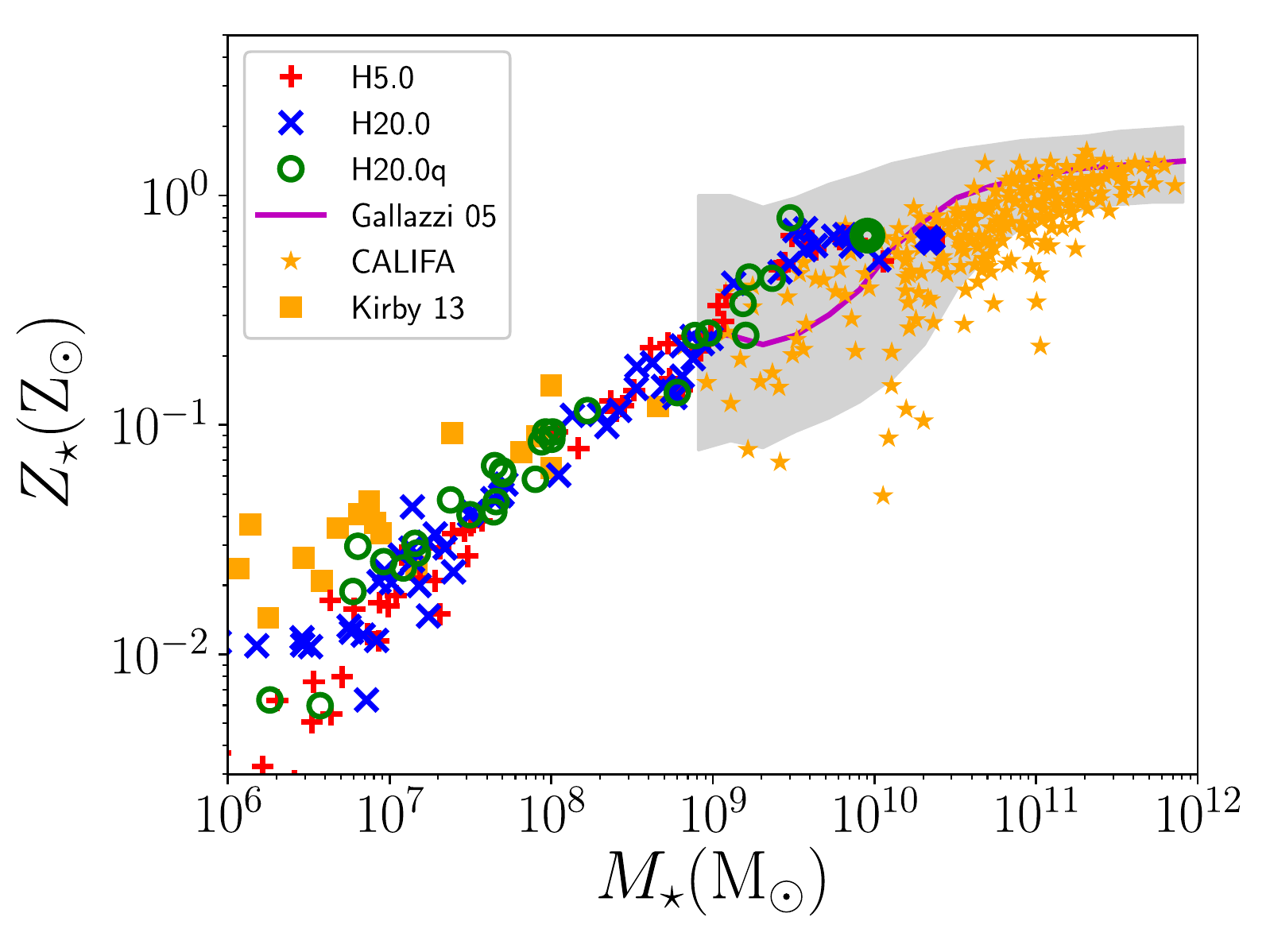}
\caption{Stellar mass--metallicity relation for our H-runs. The orange stars correspond to the data from the CALIFA survey \citep{delgado14}, while the cyan squared are taken from \citet{kirby13}. The magenta line corresponds to the median of the distribution from \citet{gallazzi05}, with the $16\%$ and the $84\%$ percentiles limiting the shaded area. The most massive galaxies in our three simulations are the thicker markers in the plot.}
\label{fig:mzr}
\end{figure}

\subsection{Disc dynamics: the stellar and baryonic Tully--Fisher relations}
We now compare the dynamic properties of our galaxies with both the stellar and baryonic Tully--Fisher (TF) relation. 
The top panel in Fig.~\ref{fig:tf} shows the stellar TF relation for our sample, compared with the data and the orthogonal fit by \citet[][AR08 hereafter]{avilareese08}, where the grey band corresponds to the $1\sigma$ scatter, and the fit by \citet[][F14 hereon]{ferreras14} for galaxies at $z>0.45$. A clear consensus on the evolution of the relation with redshift is still missing, so we do not consider any possible redshift scaling in this analysis. {We observe that
the simulation data almost perfectly lie on the relation by F14 for stellar masses above $10^8\, \msun$, except for the most massive galaxy in H20q, probably because of its excessive compactness. At low masses, instead, the low efficiency of SF leads to an increased scatter in the data and, in the stellar TF case, to a steepening of the relation. Compared to AR08, instead, our data points are slightly offset towards lower masses/higher velocities, but this could be in principle compensated if the stellar TF would evolve with redshift.}

In the bottom panel of Fig.~\ref{fig:tf}, we report the baryonic TF relation, which has been advocated to be tighter compared to the stellar counterpart. We compare here our simulations with the AR08 data only, since these are not available in F14. In this case, we find very good agreement with observational data, except for the central galaxy in H20.0q, slightly offset compared to the relation, while H20.0 and H5.0 are perfectly consistent with the data. At low masses, we do not observe anymore a steepening of the relation, but an increased scatter almost equally distributed above and below the AR08 fit.

\subsection{The stellar mass--metallicity relation}
We now compare the average stellar metallicity in our simulations with the observational data of low-redshift galaxies. We stress again that the observational data reported here are obtained with different techniques (e.g. using different apertures, weightings, etc.) which make difficult to directly compare them with simulations. However, a detailed analysis of the different caveats is beyond the scope of this work, where we are only interested into a comparison of the trend and of the metallicity range for the different galaxy masses. We show in Fig.~\ref{fig:mzr}, the values obtained for our galaxies as a function of the total stellar mass. The orange stars correspond to the data from the CALIFA survey \citep{delgado14}, while the orange squares are data from \citet{kirby13}. The magenta line corresponds to the median in \citet{gallazzi05}, with the $16\%$ and $84\%$ of the distribution delimiting the grey area. 
We note that our simulations match very well the observational data, especially at low masses. On the other hand, there is a slight increase in metallicity in the intermediate range between $10^9$ and $10^{10}\, \msun$, which is the range where our simulations show the largest deviation from the stellar to halo mass relation (see Fig.~\ref{fig:shmrelation}). The consistency between these two comparisons strengthens the idea that in this mass range the SN feedback model used starts to be ineffective in suppressing SF, resulting in overestimated stellar masses and slightly higher metallicities.

\section{The high-resolution runs: redshift evolution}
\label{sec:redshift}
In the previous section (Fig.~\ref{fig:msrelation}) we have shown that, for the most massive galaxies in our sample, moving from 5 to 20 H cm$^{-3}$ for the SF density threshold results in approximately the same total stellar mass (also noticeable from Fig.~\ref{fig:shmrelation}), but with a slightly smaller half-mass radius. This result  can be explained based on the SF prescription: a higher density threshold requires the gas to collapse further before matching the conditions for SF, but since the SFR scales as $\rho^{1.5}$, the SFR would be higher, leading to approximately the same total gas mass converted into stars.
However, very high densities are typically reached in the centre of galaxies, which translates into a slightly more compact stellar distribution. The previous section has shown that galaxy properties have a small dependence on the SF density threshold adopted. Therefore, since this effect is small, for the analysis of redshift evolution of the galaxy properties between $z=4$ and $z=0.5$, we group all the three H-runs together, averaging the properties of the full sample.

The simulation sample is here compared with the observational data and theoretical predictions. In order to make the figures clear, we limit the analysis to two mass bins with half-decade width. For the stellar to halo mass relation, we use two mass bins centred  around $10^{10}$ and $10^{11}\,\msun$  in halo mass, for the baryonic TF relation we centre around $10^8$ and $10^{10}\,\msun$ in baryonic mass and for all the other analyses we bin in stellar mass around $10^8$ and $10^{10}\,\msun$. The mass bins are kept constant at all redshift and we do not follow the histories of individual haloes.  The typical number of galaxies in the stellar mass bins is 20 ($3\times 10^7\,\msun <M_\star<3\times 10^8\,\msun$) and 5 ($3\times 10^9\,\msun <M_\star<3\times 10^{10}\,\msun$), while for the baryonic mass bins it is 10 ($3\times 10^7\,\msun <M_{\rm bar}<3\times 10^8\,\msun$) and 25 ($3\times 10^9\,\msun <M_{\rm bar}<3\times 10^{10}\,\msun$).

\begin{figure}
\centering
\includegraphics[width=0.48\textwidth]{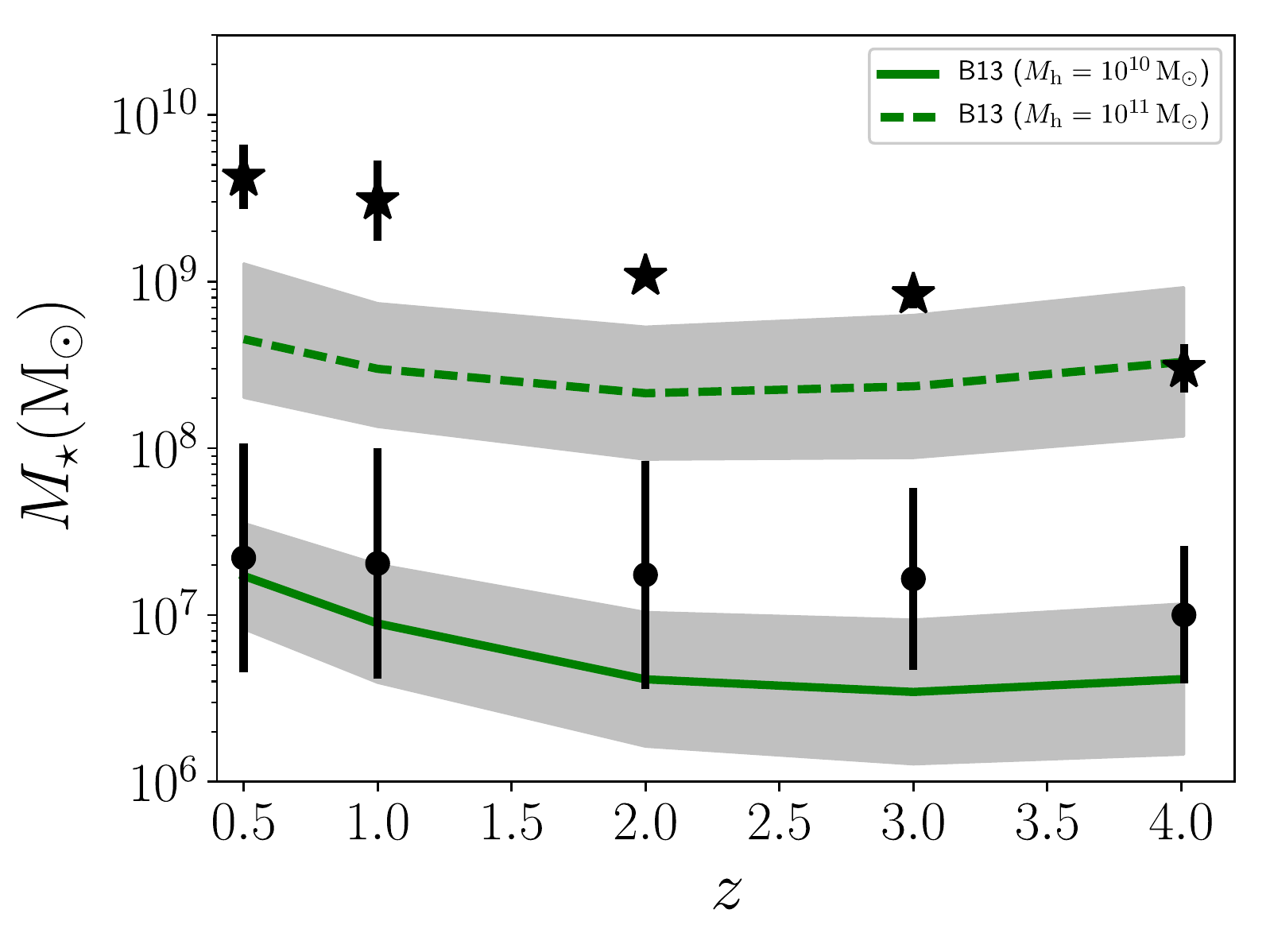}
\caption{Redshift evolution of the stellar mass--halo mass relation for our H-runs. We report the average stellar mass evolution for our galaxies from $z=4$ to $z=0.5$ as black dots/stars. We bin in halo mass, using two bins centred around $10^{10}$ (dots) and $10^{11}\,\msun$ (stars), respectively, each bin covering half a decade. The error bars correspond to $1\sigma$ uncertainty for the values in the bins. The green lines correspond to the B13 relation evaluated for both stellar mass bins at every redshift, with the shaded area identifying the $1\sigma$ uncertainty of the model.}
\label{fig:shr_z}
\end{figure}

\subsection{Stellar to halo mass relation}
Fig.~\ref{fig:shr_z} shows the evolution of the stellar mass as a function of redshift, in black. The green lines correspond to the B13 relation, with the shaded area corresponding to $1\sigma$ dispersion. For low-mass galaxies, the stellar mass is larger than the theoretical predictions, but usually within $2\sigma$. We also note that the strongest discrepancy appears between $z=2$ and $z=1$, while a better agreement is recovered at the final redshift of the simulations. On the other hand, more massive galaxies do not converge back to the predicted relation at low redshift, but stay well above it. 
\subsection{Mass--size relation}
\begin{figure}
\centering
\includegraphics[width=0.48\textwidth]{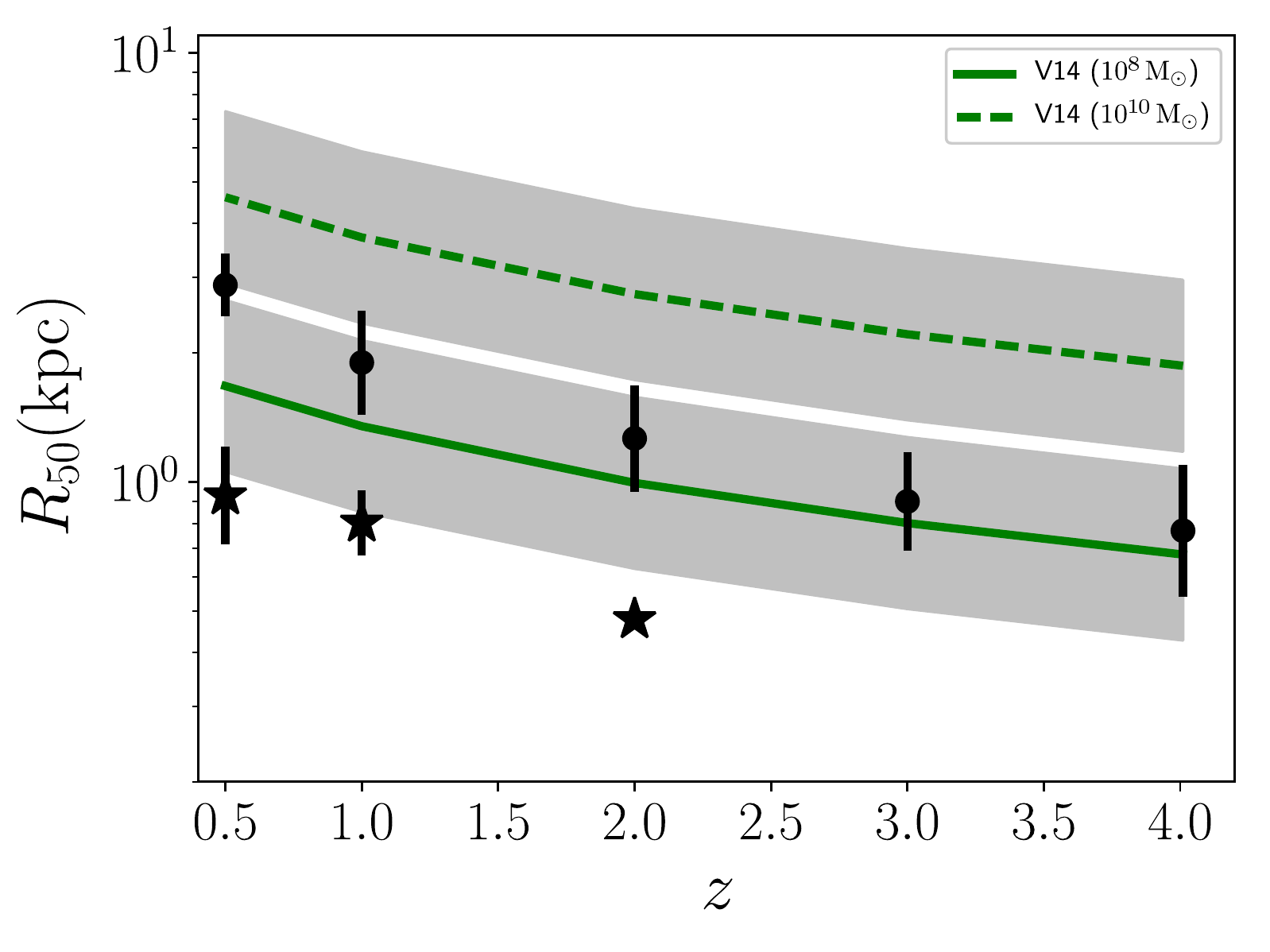}
\caption{Redshift evolution of the mass--size relation for our H-runs. We report the average distribution of our galaxies from $z=4$ to $z=0.5$ as black dots/stars. We bin in stellar mass, using two bins centred around $10^8$ (dots) and $10^{10}\,\msun$ (stars), respectively, each bin covering half a decade. The error bars correspond to $1\sigma$ uncertainty of the data in the bins. The green lines correspond to the relation for late-type galaxies from the 3D+HST+Candels survey, with the shaded area identifying the observational $1\sigma$ uncertainty.}
\label{fig:msr_z}
\end{figure}

We report in Fig.~\ref{fig:msr_z} the redshift evolution of the mass--size relation, in black, with the error bars identifying the standard deviation of the distribution. We compare our results with the data from the 3D+HST+Candels survey \citep{vanderwel14}, in green, where the shaded area corresponds to $1\sigma$ dispersion. While low-mass galaxies are reasonably consistent with the observed data, being well within $1\sigma$ except for the final redshift, the massive counterparts are always compact, with typical sizes $\lesssim$ 1 kpc, in disagreement with observations. As already mentioned, a possible explanation is that the SN feedback is not effective enough in suppressing SF, especially in the centre where the density is usually high, thus resulting in too many stars concentrated within the central kpc.
\subsection{Stellar and baryonic TF}
\begin{figure}
\centering
\includegraphics[width=0.48\textwidth]{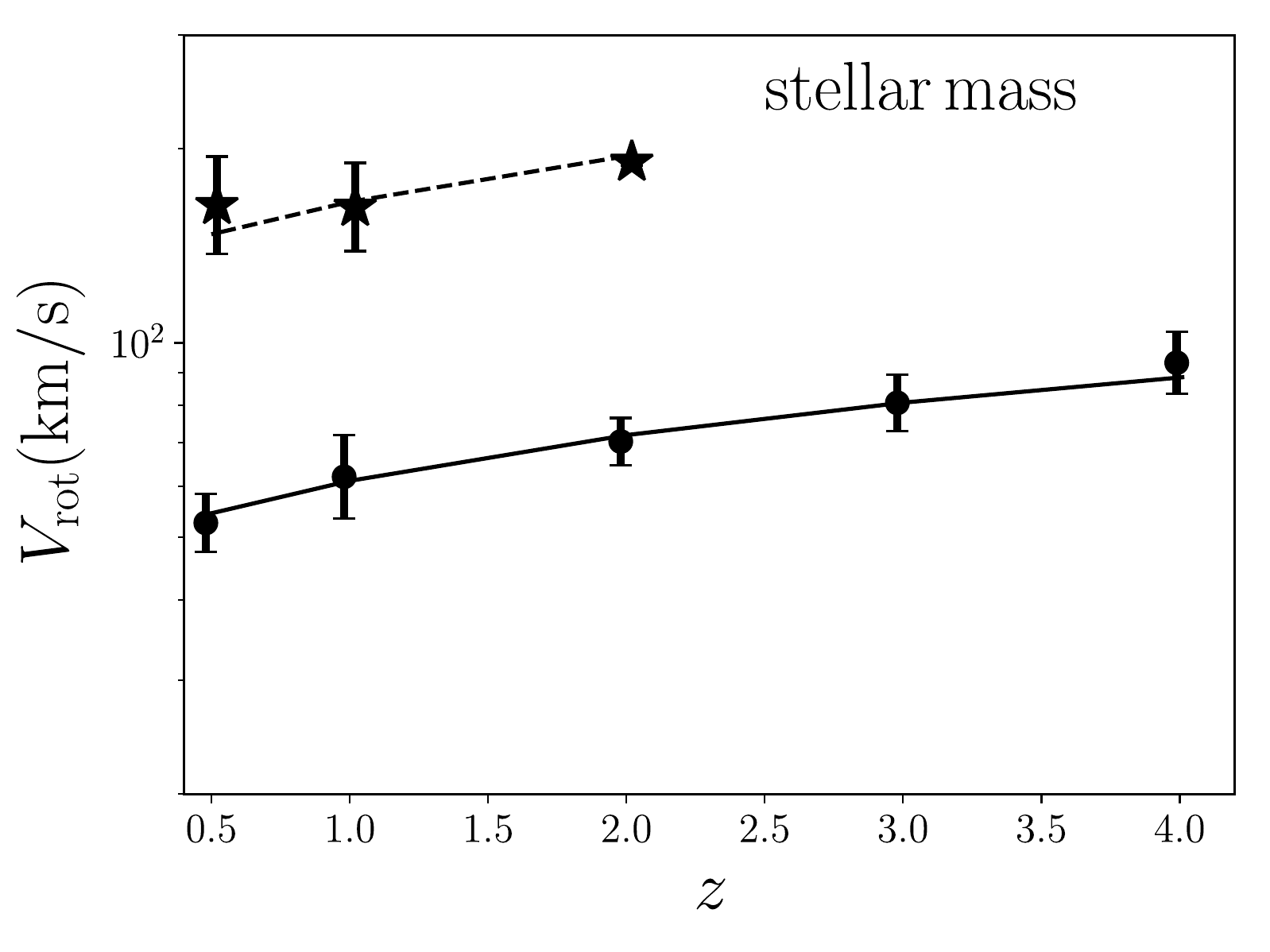}
\includegraphics[width=0.48\textwidth]{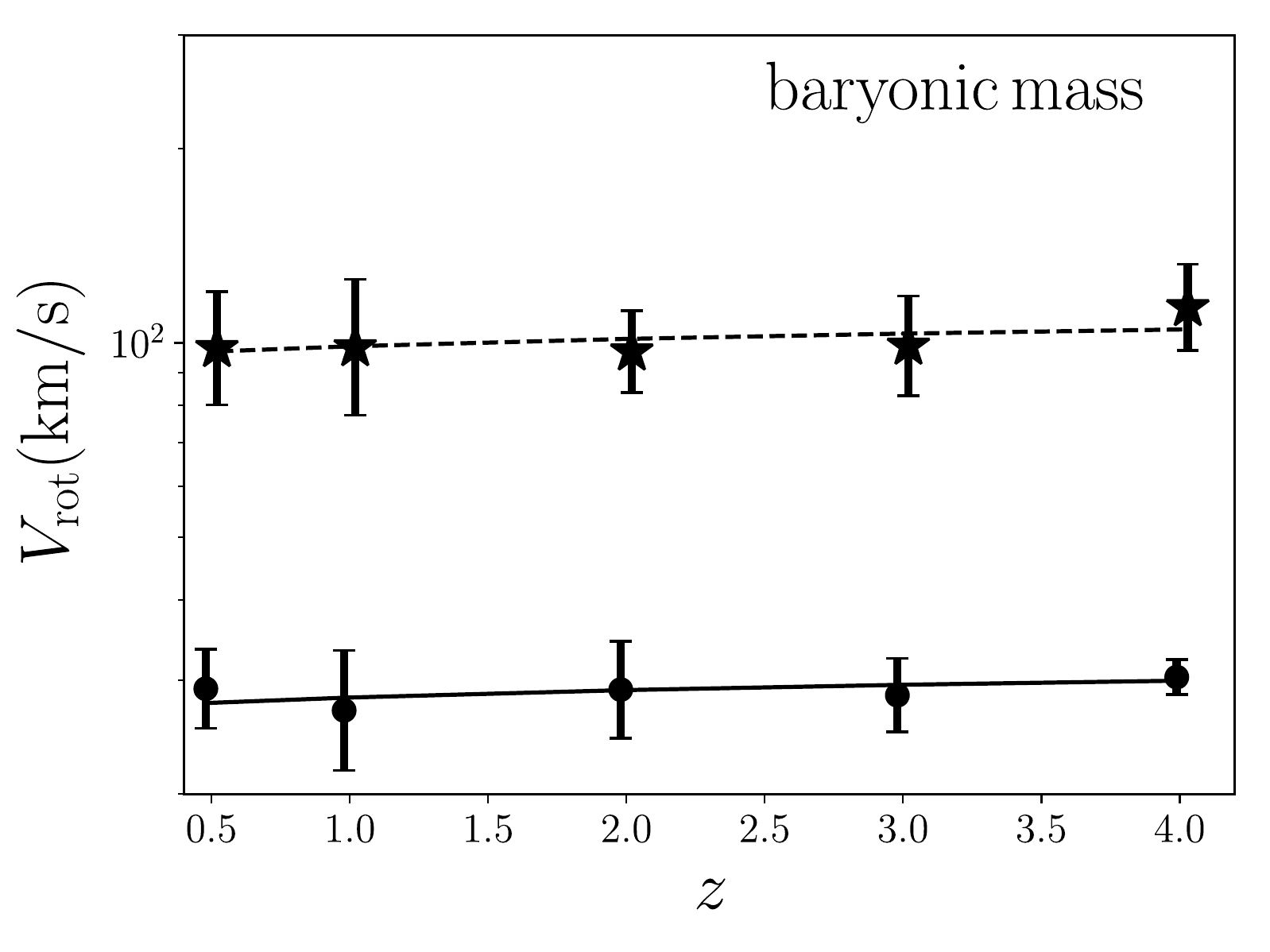}
\caption{Redshift evolution of the TF relations for our H-runs. We report the average distribution of the rotational velocity $V_{\rm rot}$ for the simulated galaxies from $z=4$ to $z=0.5$ as black dots/stars. {The only difference in the two panels is the binning.} The error bars correspond to the $1\sigma$ uncertainty in the bin. Top panel: rotational velocity in two stellar mass bins around $10^8$ (dots) and $10^{10}\,\msun$ (stars), respectively, each bin covering half a decade. Bottom panel: same as top panel, but binned in total baryonic mass, with two bins centred on $10^8$ (dots) and $10^{10}\,\msun$ (stars), respectively, half a decade wide. The black lines (solid for $10^8\,\msun$ and dashed for $10^{10}\,\msun$) correspond to the best fits to the data using the power-law model defined in equation~(\ref{eq:vrotz}).}
\label{fig:tf_z}
\end{figure}

The top panel in Fig.~\ref{fig:tf_z} shows the evolution of the galaxy rotational velocity $V_{\rm rot}$ (defined in \S\ref{sec:limits}) as a function of redshift, in black, with the error bars identifying the standard deviation of the distribution. The data points correspond to the average value for galaxies in two mass bins, centred around  $10^8$ (dots) and $10^{10}\,\msun$ (stars) in stellar mass. We do not follow the galaxy history, but we keep the mass bins fixed at different redshifts. The black lines in the plot correspond to the best fit to the data points using a power-law model defined as
\begin{equation}
V_{\rm rot} (M_x,z) = V_{\rm rot,0}(M_x/M_{x,0})^\alpha(1+z)^\beta,
\label{eq:vrotz}
\end{equation}
where $V_{\rm rot,0}$ is the rotational velocity at $z=0$ for the reference mass $M_{x,0}$, $M_x$ is the stellar/baryonic mass of the bin, $\alpha$ is the mass scaling exponent (expected to be $\sim 0.25$) and $\beta$ is the redshift scaling exponent. The best fit to the data gives $M_{x,0}= 10^{9.02}\,\msun, V_{\rm rot,0}=10^{1.89}{\rm\, km/s}, \alpha= 0.22$ and $\beta=0.40$.
In the bottom panel of Fig.~\ref{fig:tf_z}, instead, we plot the galaxy rotational velocity for galaxies with a total baryonic mass of $10^8$ (dots) and $10^{10}\,\msun$ (stars), using the same approach as for the stellar counterpart. Also in this case, the black lines are the best fits to the data points with equation~(\ref{eq:vrotz}), with parameters $M_{x,0}= 10^{9.08}\,\msun, V_{\rm rot,0}=10^{1.72}{\rm\, km/s}, \alpha= 0.27$ and $\beta=0.06$. The baryonic relation shows only a mild evolution, reasonably consistent with no evolution at all, while a clear trend can be observed for the stellar counterpart.
If we consider the trend we find here for the stellar TF relation and rescale the data points from $z=0.5$ to $z=0$ in Fig.~\ref{fig:tf} (top panel) by a factor $(1+z)^\beta$, with $\beta=0.4$, we get a better agreement with the AR08 relation discussed in the previous section.

\subsection{Stellar mass--metallicity relation}
\begin{figure}
\centering
\includegraphics[width=0.48\textwidth]{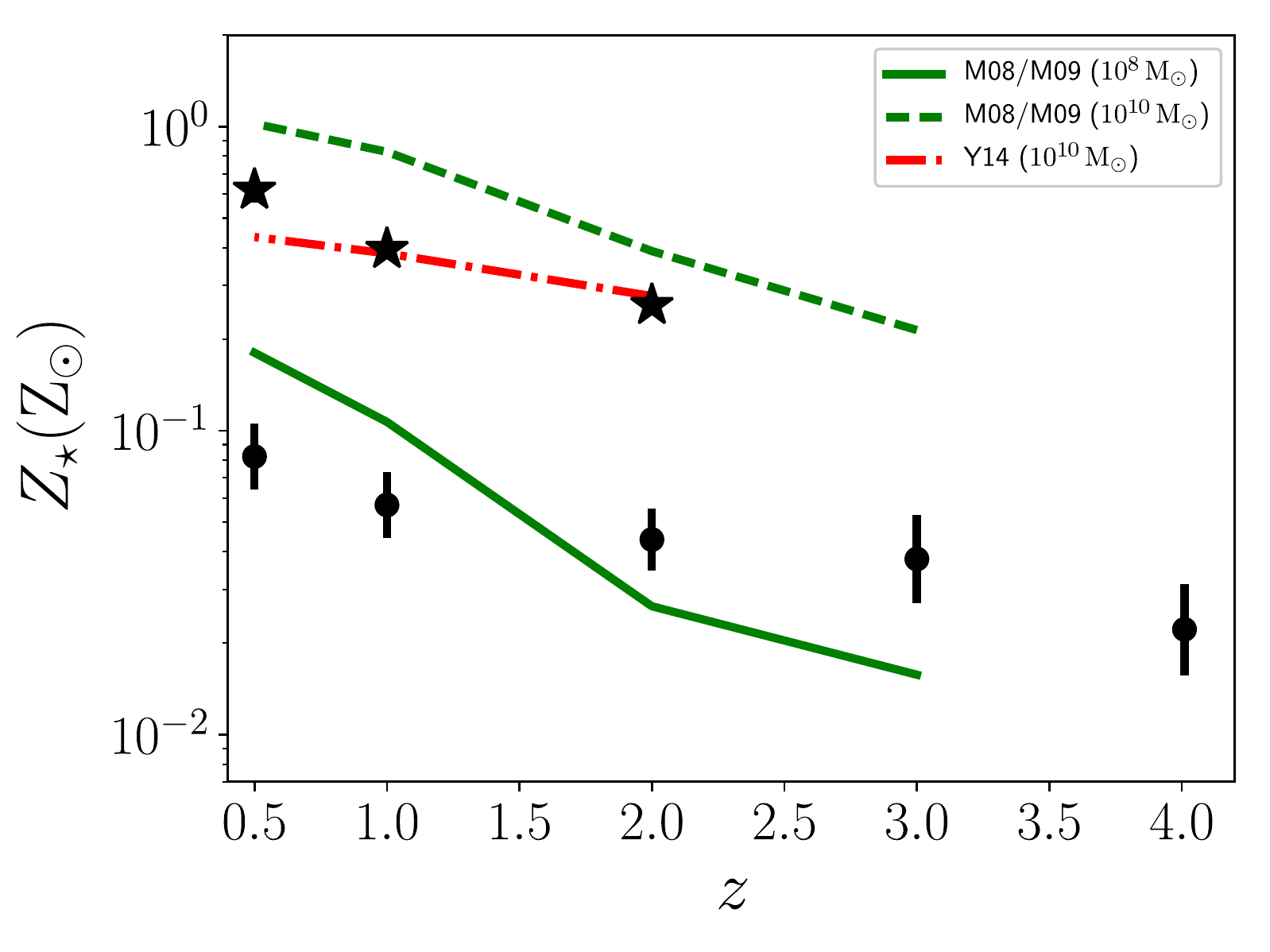}
\caption{Redshift evolution of the stellar mass--metallicity relation for our H-runs. We plot the average metallicity of the simulated galaxies in solar units \citep[where $Z_\odot = 0.02$ is the solar metal fraction,][]{anders89} (as black dots/stars), compared to the data by \citet{maiolino08,mannucci09} (solid/dashed green lines) and \citet{yabe14} (dot--dashed red line) at $10^8\,\msun$ and $10^{10}\,\msun$, respectively. The error bars associated with the simulated data correspond to $1\sigma$ uncertainty of the data in the bins.}
\label{fig:mzr_z}
\end{figure}

Fig.~\ref{fig:mzr_z} shows the evolution of the average stellar metallicity as a function of redshift, in black, for our simulated galaxies. We use two stellar mass bins, $3\times 10^7\,\msun <M_\star<3\times 10^8\,\msun$ and $3\times 10^9\,\msun <M_\star<3\times 10^{10}\,\msun$, as in the previous analysis. The error bars identify the standard deviation of the distribution. We compare the simulated galaxies with the data by \citet[][M08 hereafter]{maiolino08} and \citet[][M09 hereafter]{mannucci09}, plotted as solid and dashed green lines, up to $z\sim 3$, and by \citet[][Y13]{yabe14}, as a dot--dashed red line, up to $z\sim 1.4$. We compute the value reported in the plot using the fitting functions provided by the authors for $10^8$ and $10^{10}\,\msun$, respectively (equation~(2) in \citet{maiolino08} and equation~(3) in \citet{yabe14}), with the appropriate coefficients for the different epochs. For this comparison, we assume a solar metallicity based on the oxygen abundance defined as $\ln \tilde{Z}_\odot = 12+\ln(O/H) = 8.91$ \citep{anders89}, to be consistent with the comparison in the previous section. For low-mass galaxies, we observe a clear trend consistent with the observations, although the observational constraints are not very tight. For the most massive galaxies, instead, we find a slightly lower metallicity compared to M08/M09, consistent with Y13.

\section{The low-resolution runs: effect of the  choice of parameters}
\label{sec:lowres}
In this section, we describe the results of the low-resolution simulations at $z=0.5$ (to be consistent with the latest redshift of the H-runs), analysing the differences in the galaxy properties depending on the parameter choice, namely the SF density threshold (L0.2, L1.0 and L5.0) and the SF efficiency (L1.0low). Since we are interested here into a galaxy by galaxy comparison, a clear matching of the galaxies in the different runs would be more difficult for the low-mass galaxies compared to the high mass ones. Therefore, for this specific comparison, we limit the analysis to the five most massive galaxies in the zoom-in region identified by \textsc{ahf} (G1--5), where the most massive one corresponds to the main halo in the box. The galaxies are matched across the different runs via a mass and position criterion, looking for galaxies with almost identical halo mass and with the closest halo centre of mass among the different runs.
\begin{figure}
\centering
\includegraphics[width=0.48\textwidth]{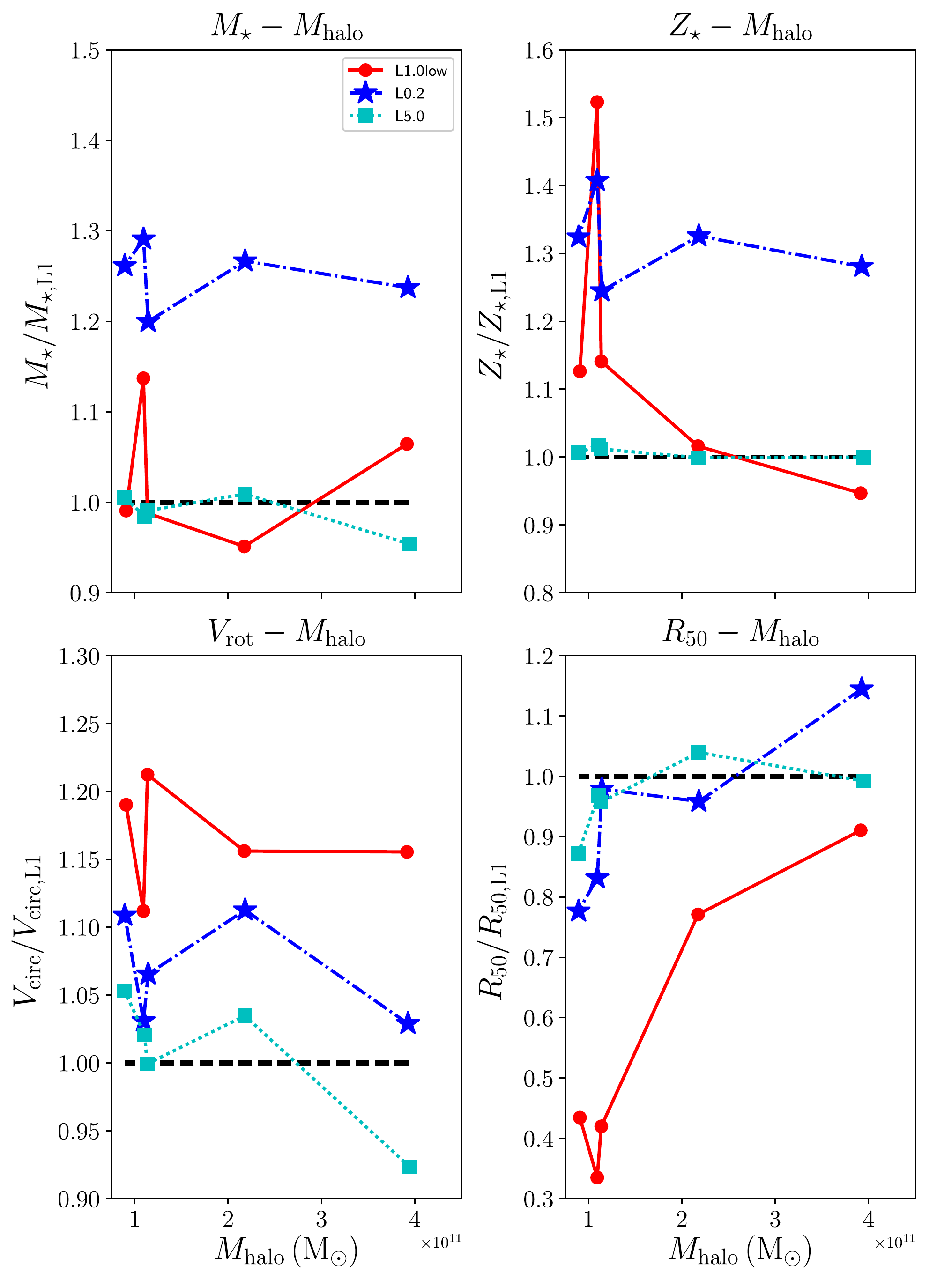}
\caption{Comparison of the galaxy properties for the five most massive haloes in our L-runs. We consider L1.0 as the reference run for this comparison (black dashed line). L0.2 is the blue dash--dotted line (with the blue stars), L1.0low is the red dashed line (with the red circles) and L5.0 is plotted with a cyan dotted line (with the cyan squares). The top-left panel shows the total stellar mass, the top-right one the average stellar metallicity, the bottom-left one the rotational velocity and the bottom-right.}
\label{fig:propcomp}
\end{figure}

We compare here the main properties of the selected galaxy sample, stellar mass, average stellar metallicity, rotational velocity and effective size, using the L1.0 run as reference. All the quantities are estimated following the same procedure described in \S\ref{sec:results}.

\begin{figure*}
\centering
\includegraphics[width=0.75\textwidth]{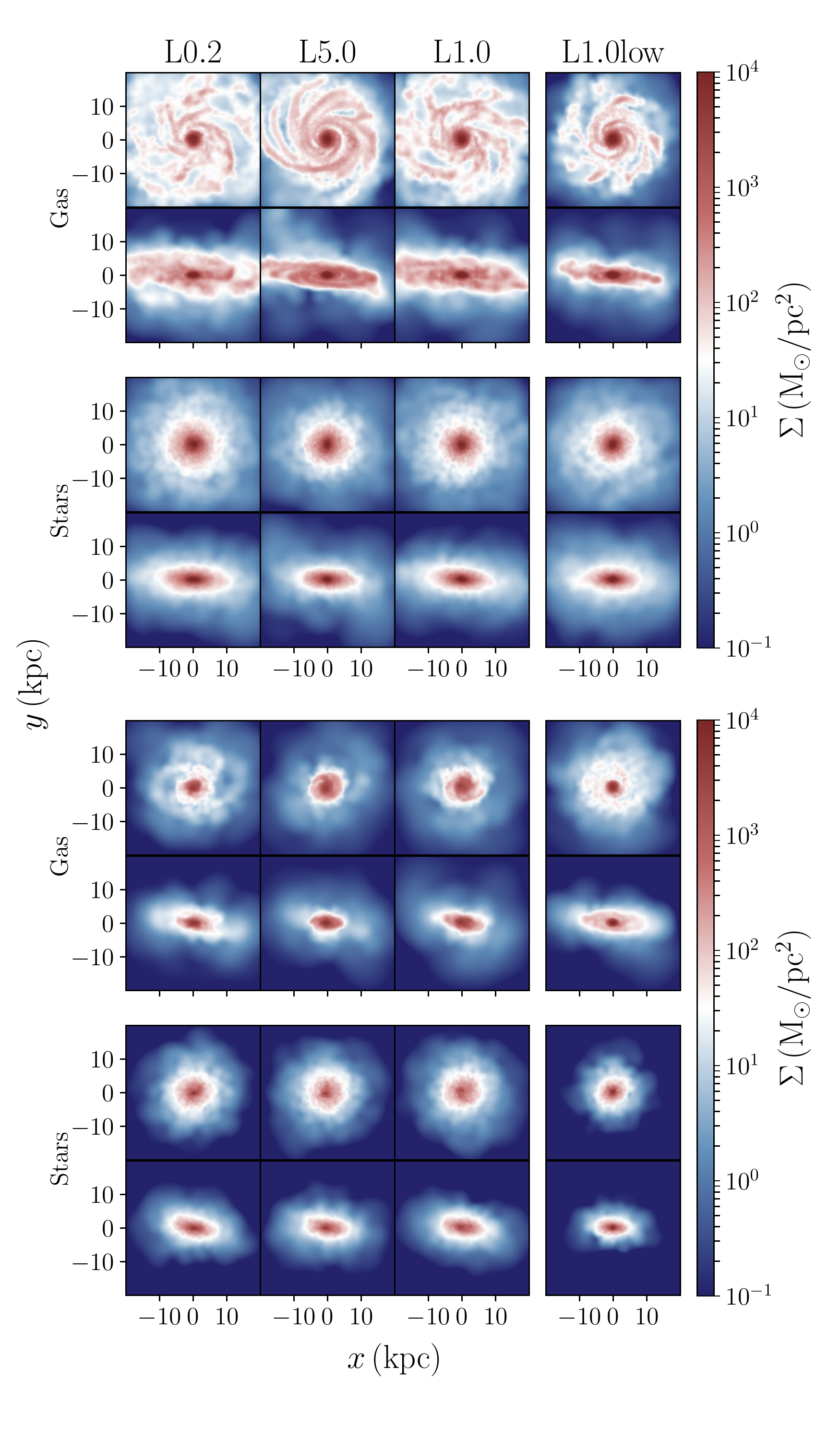}
\caption{Projected density maps for G1 and G5 in our L-runs. The first column corresponds to L0.2, the second to L5.0,the third to L1.0 and the right-most one to L1.0low. The first 4 rows show both gaseous (top) and stellar (bottom) components for G1 and the last 4 rows are the same, but for G5. {For both galaxies we show the face-on (rows 1-3-5-7) and the edge-on views (rows 2-4-6-8).}}
\label{fig:lowresmap}
\end{figure*}

\begin{figure*}
\centering
\includegraphics[width=0.85\textwidth]{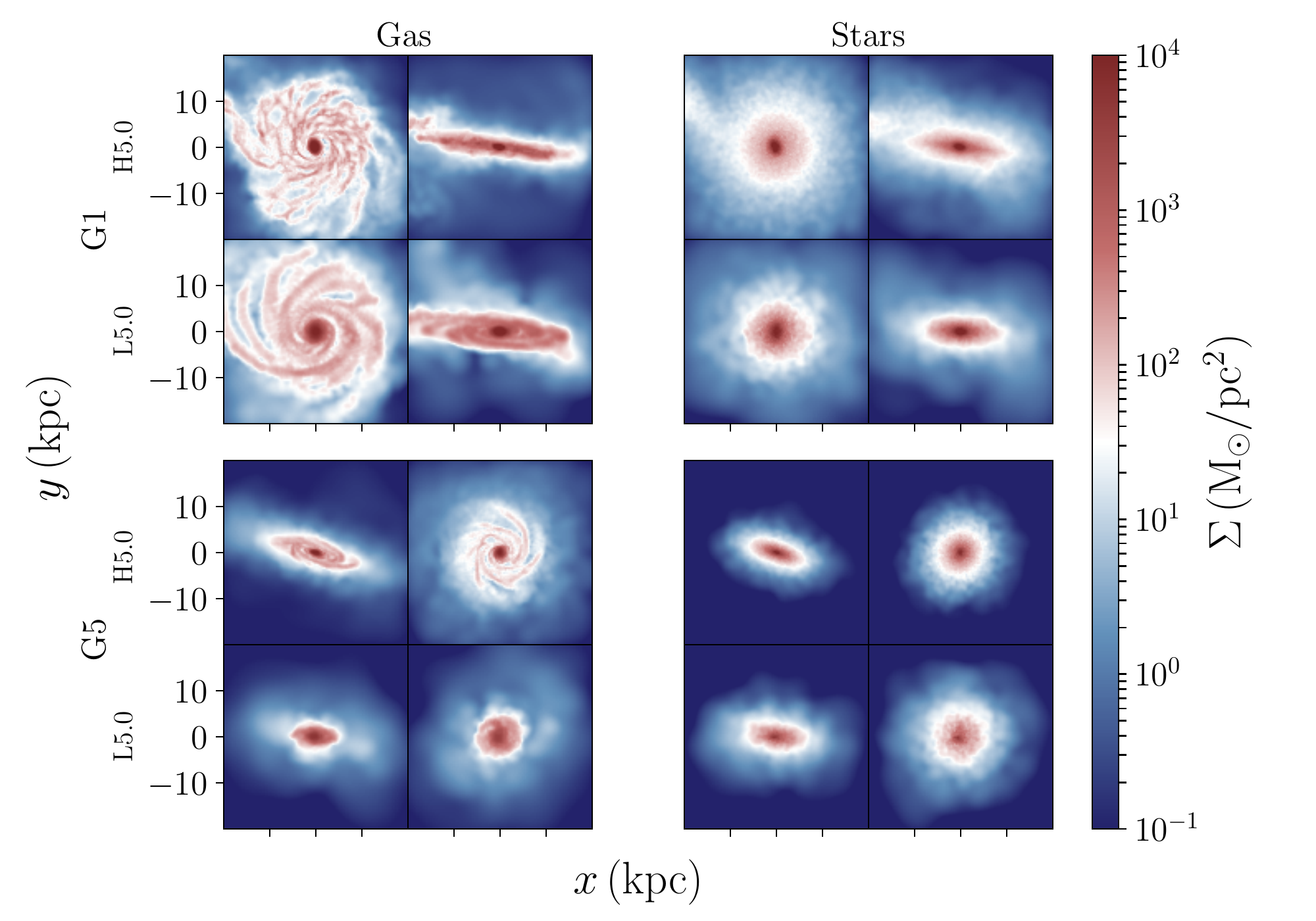}
\caption{Projected density maps for G1 and G5 in H5.0 and L5.0.  Rows 1-2 show the gaseous (left-hand panels) and the stellar (right-hand panels) component for G1 in the two runs. Rows 3-4 are the same plots for G5.}
\label{fig:highlowmap}
\end{figure*}

Fig.~\ref{fig:propcomp} shows the comparison among the different runs. 
In the top-left panel, we report the variation of the stellar mass with respect to L1.0 (represented by a dashed black line). Simulation L0.2 overpredicts the stellar mass by roughly $25\%$, because of the very low density threshold, while L5.0 is almost identical to L1.0. Run L1low shows a higher variability between $75\%$ and $115\%$, but overall is consistent with L1.0, showing that the SF efficiency does not change much the total amount of stars formed.  

In the top-right panel, we show the average stellar metallicity relative to L1.0. Also in this case, L5.0 is consistent with L1.0, with very small differences, while L0.2 has higher metallicity $\sim 1.3$ times that of L1.0 at fixed halo mass, corresponding to the higher stellar mass formed. Run L1.0low exhibits a strong scatter, with a decline towards the galaxy masses, again simply following the trend given by the stellar mass. 

In the bottom-right panel, we plot the effective size of the galaxy, defined as the radius enclosing $50\%$ of the galaxy stellar mass (see \S\ref{sec:results}). The ratio of the size to that of L1.0 increases at higher halo mass in all the runs, with clear positive trend visible for L0.2, while L5.0 settles around 1 with a scatter of $\sim 10\%$. If the SF density threshold is too low, it affects both how many stars are formed, {\it and} where they form. The relative size of run L1.0low is always smaller,  between 0.4 and 0.9 that of L1.0. The efficiency of SF therefore does not affect the total stellar mass in the galaxy, {\it but} it has an impact on where they form and thus on the typical galaxy size \citep{agertz11}.

Finally, in the bottom-left panel, we report the rotational velocity of our galaxies. In this case, L5.0 is the closest one to L1.0, but showing a negative trend with increasing galaxy mass. L0.2, instead, has a velocity in excess of $5-10\%$. L1.0low shows the largest discrepancy,  $\sim 15-20\%$ reflecting that both the stellar mass and the distribution differ with respect to L1.0. 

In conclusion, L0.2 and L1.0 show instead clear differences in the galaxy properties, while L5.0 and L1.0 behave in a similar way, suggesting that the density threshold, once sufficiently high, does not play a significant role in determining the total stellar mass. L1.0low also shows noticeable differences with respect to L1.0, highlighting, in particular, that the SF efficiency affects the galaxy stellar distribution more than the SF threshold.

The quantitative trends described above can also be appreciated from a visual comparison of the gas and stellar projected surface densities for G1 (rows 1--4) and G5 (rows 5--8), shown in Fig.~\ref{fig:lowresmap}.  Rows 1--2 and 5--6 show the gas surface density along the z- (top panels) and y-axes (bottom panels) in a box of 40 kpc side centred on the galaxy centre of mass, while the remaining rows correspond to the same maps for the stellar component. From left to right, we vary the SF density threshold (first 3 columns) and the SF efficiency (right-most column). The gas projection is computed using \textsc{splash} \citep{splash07}, by integrating the density field along the line of sight with the smoothing kernel. For the stellar component, not carrying the smoothing length information, we compute the density field using a kd-tree, defining the kernel size as the spherical radius encompassing exactly 32 neighbour stellar particles and then project it as done by \textsc{splash}.

For G1, we observe a well-defined gaseous disc, with clearly visible spiral arms. In L0.2, because of the lower SF density threshold, the SFR is slightly higher and the SF more diffuse, with SNe having a stronger effect in perturbing the disc. On the other hand, the spiral arms are clearly defined in L5.0, where the higher SF density threshold prevents SF from occurring  in low gas density regions. The gas concentration in the nucleus and the gaseous disc size are comparable in all of these three runs. L1.0low, instead, shows a more compact gaseous disc, with a slightly denser gas in the nucleus, because of the globally lower SF efficiency which prevents SF throughout the disc, allowing for stronger inflows towards the centre. For the stellar component, we observe a similar behaviour, with a slightly more extended and massive stellar disc in L0.2 compared to the other runs (consistent with Fig.~\ref{fig:propcomp}, top-left panel) and a somewhat more compact stellar disc in L1.0low. L1.0 and L5.0, instead, are more similar, reflecting the quantitative results described above.

For G5, we observe a much less well-defined disc, except for L1.0low, where the low SF efficiency allows the gas to cool and settle down without being heavily perturbed by SNe. The gas structure in L0.2, L1.0 and L5.0 is similar, with a moderately more massive concentration in L5.0 compatible with a less efficient SF at larger radii, which allows more gas to collect into the central region of the galaxy, as discussed in \S\ref{sec:results}. For the stellar component, we observe a similar behaviour. The stellar distribution in L0.2, L1.0 and L5.0 does not differ significantly, while L1low reveals a denser and more compact structure. 

Finally, we show in Fig.~\ref{fig:highlowmap}, a comparison between H5.0 and L5.0 for G1 and G5, to highlight the differences obtained by increasing the mass and spatial resolution at fixed SF density threshold. In both galaxies, the gaseous disc is better resolved in H5.0 and shows more detailed features, but the main properties are preserved. We observe the same behaviour for the stellar component. The stellar and gaseous overdensity in the centre, responsible for the observed compactness, is present in both runs, but with a sharper density gradient in the H-run. 

\section{Discussion and Conclusions}
\label{sec:conclusions}
We have presented here a suite of zoom-in cosmological simulations performed with the new cosmological code \textsc{gizmo} \citep{hopkins15}, using the MFM scheme for hydrodynamics.
The code has demonstrated very good performance on the standard benchmark tests, but it has not been tested yet with cosmological simulations using  delayed cooling SN feedback. In this study, we implemented these models in the code, as described in \S\ref{sec:setup}, and investigated the  ability of the code, coupled with them, to reproduce observed scaling relations for galaxies. We ran a suite of low- and high-resolution zoom-in simulations with a high-resolution region centred around a halo with few $10^{11}\,\msun$ at $z=0$. In the four low-resolution simulations, we varied the SF density threshold and the SF efficiency, while in the three high-resolution simulations we considered only two different SF density thresholds and a different merger history.

 For the high-resolution runs, we compared the galaxy properties at the final redshift ($z=0.5$) with the observed scaling relations and theoretical models (at $0<z<0.5$). 
We found that we can reproduce the main properties reasonably well, with the exception of the mass--size relation for the most massive galaxies, which are too compact compared to the population of late-type galaxies that they should belong to and the stellar to halo mass relation when compared to the empirical model by B13. 

We consider now different possible reasons for the excessive compactness. First, the SN feedback model, as implemented in the code, could be ineffective in the dense regions of the most massive galaxies. Because of the energy deposition scheme for SNe, the number of neighbours in the kernel can exceed the desired number 32 when the minimum smoothing length is reached, resulting in a smaller energy injection per particle.Gas particles are heated to lower temperatures, hence less effective in expelling the gas from the centre. Moreover, when the shutoff time for gas cooling has passed, these `warm' gas particles would easily match again the SF criteria, resulting in a new SF burst in the centre. More refined models for SN feedback fare better in this regime \citep{oppenheimer10}.
Secondly, because of the rather low resolution in our simulations, where the most massive galaxies are modelled with no more than $\sim 10^5$ gas particles, our simulations could suffer some spurious transfer of angular momentum driving the gas towards the centre, favouring the formation of a more massive bulge. This issue has been found to plague SPH codes, where the artificial viscosity and the smoothing procedure on close-by particles with highly different temperatures can lead to numerical losses of angular momentum \citep{mayer08,torrey12}. The new hydrodynamics scheme used here, able to accurately resolve shocks and fluid mixing, should reduce this effect, but we would need higher resolution runs to eventually exclude it as possibly responsible for the observed results, since a similar behaviour has been observed as well in our low-resolution runs.
Finally, as discussed in \citet{dubois13}, the lack of a central BH in massive galaxies can lead to the formation of too compact galaxies. Although the simulated galaxies are at the low-mass end of the mass distribution, these BHs should be small, but nevertheless they could play a crucial role in reshuffling the gas in the central few hundred pc and suppressing the SF in the centre. However, a detailed investigation of this issue and the inclusion of a BH population are beyond the scope of the present study, and will be discussed in a future paper. 

As for the stellar to halo mass relation, we found that we typically overpredict the total stellar mass for the most massive galaxies in the sample. Although our results seem to be in agreement with observed nearby galaxies, the already mentioned ineffectiveness of the SN feedback in the centre of the massive haloes we simulated could also play a role. For instance, we get up to a factor of 10 more stars than B13 for haloes around $10^{11}\,\msun$, with a discrepancy increasing with time.

We also discussed the redshift evolution of the analysed galaxy properties, finding good agreement with observational data, where available. The only exceptions are the stellar to halo mass relation, where we overpredict the value compared to the theoretical models at low redshift, and, again, the mass--size relation for the most massive haloes in our runs.

 For the low-resolution runs, instead, we compared the main galaxy properties (stellar mass, rotational velocity, average stellar metallicity and half-mass size) for the five most massive haloes in our low-resolution runs, to assess the effect of the parameter choices . We found that a lower SF efficiency tends to produce more compact galaxies, while a lower SF density threshold results in a significant overprediction of the total stellar mass and metallicity. On the other hand, the two highest values of the SF density threshold show very good agreement, suggesting that above a certain density threshold, the evolution becomes much less dependent on this parameter choice. 

The code, coupled with this simple model for SF and SN feedback, has proved to be in reasonable agreement with observations, within the limits of the methods used to extract the right information from the simulations. The models used here are not the true answer for galaxy formation, but they are a good starting point to study galaxy formation, especially when resolution  is not sufficient to resolve in detail the multiphase medium in the galaxies, which is the case for large cosmological volumes.

\section*{Acknowledgments}
We thank the anonymous referee for constructive comments that helped us to improve the quality of the paper. We acknowledge support from the European Research Council (Project No. 267117, `DARK', AL, JS; Project no. 614199, `BLACK', AL, MV). We thank P.F. Hopkins for useful discussion and suggestions.  This work was granted access to the High Performance Computing resources of TGCC  and CINES under the allocation x2016046955 made by GENCI, and it has made use of the Horizon Cluster, funded by Institut d'Astrophysique de Paris, for the analysis of the simulation results.
\bibliographystyle{mnras}
\bibliography{./Biblio}

\begin{thebibliography}{}
\makeatletter
\relax
\def\mn@urlcharsother{\let\do\@makeother \do\$\do\&\do\#\do\^\do\_\do\%\do\~}
\def\mn@doi{\begingroup\mn@urlcharsother \@ifnextchar [ {\mn@doi@}
  {\mn@doi@[]}}
\def\mn@doi@[#1]#2{\def\@tempa{#1}\ifx\@tempa\@empty \href
  {http://dx.doi.org/#2} {doi:#2}\else \href {http://dx.doi.org/#2} {#1}\fi
  \endgroup}
\def\mn@eprint#1#2{\mn@eprint@#1:#2::\@nil}
\def\mn@eprint@arXiv#1{\href {http://arxiv.org/abs/#1} {{\tt arXiv:#1}}}
\def\mn@eprint@dblp#1{\href {http://dblp.uni-trier.de/rec/bibtex/#1.xml}
  {dblp:#1}}
\def\mn@eprint@#1:#2:#3:#4\@nil{\def\@tempa {#1}\def\@tempb {#2}\def\@tempc
  {#3}\ifx \@tempc \@empty \let \@tempc \@tempb \let \@tempb \@tempa \fi \ifx
  \@tempb \@empty \def\@tempb {arXiv}\fi \@ifundefined
  {mn@eprint@\@tempb}{\@tempb:\@tempc}{\expandafter \expandafter \csname
  mn@eprint@\@tempb\endcsname \expandafter{\@tempc}}}

\bibitem[\protect\citeauthoryear{{Agertz} et~al.,}{{Agertz}
  et~al.}{2007}]{agertz07}
{Agertz} O.,  et~al., 2007, \mn@doi [\mnras]
  {10.1111/j.1365-2966.2007.12183.x}, \href
  {http://adsabs.harvard.edu/abs/2007MNRAS.380..963A} {380, 963}

\bibitem[\protect\citeauthoryear{{Agertz}, {Teyssier}  \& {Moore}}{{Agertz}
  et~al.}{2011}]{agertz11}
{Agertz} O.,  {Teyssier} R.,   {Moore} B.,  2011, \mn@doi [\mnras]
  {10.1111/j.1365-2966.2010.17530.x}, \href
  {http://adsabs.harvard.edu/abs/2011MNRAS.410.1391A} {410, 1391}

\bibitem[\protect\citeauthoryear{{Agertz}, {Kravtsov}, {Leitner}  \&
  {Gnedin}}{{Agertz} et~al.}{2013}]{agertz13}
{Agertz} O.,  {Kravtsov} A.~V.,  {Leitner} S.~N.,   {Gnedin} N.~Y.,  2013,
  \mn@doi [\apj] {10.1088/0004-637X/770/1/25}, \href
  {http://adsabs.harvard.edu/abs/2013ApJ...770...25A} {770, 25}

\bibitem[\protect\citeauthoryear{{Anders} \& {Grevesse}}{{Anders} \&
  {Grevesse}}{1989}]{anders89}
{Anders} E.,  {Grevesse} N.,  1989, \mn@doi [\gca]
  {10.1016/0016-7037(89)90286-X}, \href
  {http://adsabs.harvard.edu/abs/1989GeCoA..53..197A} {53, 197}

\bibitem[\protect\citeauthoryear{{Aumer}, {White}, {Naab}  \&
  {Scannapieco}}{{Aumer} et~al.}{2013}]{aumer13}
{Aumer} M.,  {White} S.~D.~M.,  {Naab} T.,   {Scannapieco} C.,  2013, \mn@doi
  [\mnras] {10.1093/mnras/stt1230}, \href
  {http://adsabs.harvard.edu/abs/2013MNRAS.434.3142A} {434, 3142}

\bibitem[\protect\citeauthoryear{{Avila-Reese}, {Zavala}, {Firmani}  \&
  {Hern{\'a}ndez-Toledo}}{{Avila-Reese} et~al.}{2008}]{avilareese08}
{Avila-Reese} V.,  {Zavala} J.,  {Firmani} C.,   {Hern{\'a}ndez-Toledo} H.~M.,
  2008, \mn@doi [\aj] {10.1088/0004-6256/136/3/1340}, \href
  {http://adsabs.harvard.edu/abs/2008AJ....136.1340A} {136, 1340}

\bibitem[\protect\citeauthoryear{{Beasley}, {Romanowsky}, {Pota}, {Navarro},
  {Martinez Delgado}, {Neyer}  \& {Deich}}{{Beasley} et~al.}{2016}]{beasley16}
{Beasley} M.~A.,  {Romanowsky} A.~J.,  {Pota} V.,  {Navarro} I.~M.,  {Martinez
  Delgado} D.,  {Neyer} F.,   {Deich} A.~L.,  2016, \mn@doi [\apjl]
  {10.3847/2041-8205/819/2/L20}, \href
  {http://adsabs.harvard.edu/abs/2016ApJ...819L..20B} {819, L20}

\bibitem[\protect\citeauthoryear{{Behroozi}, {Wechsler}  \&
  {Conroy}}{{Behroozi} et~al.}{2013}]{behroozi13c}
{Behroozi} P.~S.,  {Wechsler} R.~H.,   {Conroy} C.,  2013, \mn@doi [\apj]
  {10.1088/0004-637X/770/1/57}, \href
  {http://adsabs.harvard.edu/abs/2013ApJ...770...57B} {770, 57}

\bibitem[\protect\citeauthoryear{{Bottrell}, {Torrey}, {Simard}  \&
  {Ellison}}{{Bottrell} et~al.}{2017}]{bottrell17}
{Bottrell} C.,  {Torrey} P.,  {Simard} L.,   {Ellison} S.~L.,  2017, \mn@doi
  [\mnras] {10.1093/mnras/stx276}, \href
  {http://adsabs.harvard.edu/abs/2017MNRAS.467.2879B} {467, 2879}

\bibitem[\protect\citeauthoryear{{Bryan} et~al.,}{{Bryan}
  et~al.}{2014}]{bryan14}
{Bryan} G.~L.,  et~al., 2014, \mn@doi [\apjs] {10.1088/0067-0049/211/2/19},
  \href {http://adsabs.harvard.edu/abs/2014ApJS..211...19B} {211, 19}

\bibitem[\protect\citeauthoryear{{Chabrier}}{{Chabrier}}{2003}]{chabrier03}
{Chabrier} G.,  2003, \mn@doi [\pasp] {10.1086/376392}, \href
  {http://adsabs.harvard.edu/abs/2003PASP..115..763C} {115, 763}

\bibitem[\protect\citeauthoryear{{Creasey}, {Theuns}  \& {Bower}}{{Creasey}
  et~al.}{2013}]{creasey13}
{Creasey} P.,  {Theuns} T.,   {Bower} R.~G.,  2013, \mn@doi [\mnras]
  {10.1093/mnras/sts439}, \href
  {http://adsabs.harvard.edu/abs/2013MNRAS.429.1922C} {429, 1922}

\bibitem[\protect\citeauthoryear{{Dav{\'e}}, {Thompson}  \&
  {Hopkins}}{{Dav{\'e}} et~al.}{2016}]{dave16}
{Dav{\'e}} R.,  {Thompson} R.,   {Hopkins} P.~F.,  2016, \mn@doi [\mnras]
  {10.1093/mnras/stw1862}, \href
  {http://adsabs.harvard.edu/abs/2016MNRAS.462.3265D} {462, 3265}

\bibitem[\protect\citeauthoryear{{Dav{\'e}}, {Rafieferantsoa}, {Thompson}  \&
  {Hopkins}}{{Dav{\'e}} et~al.}{2017}]{dave17}
{Dav{\'e}} R.,  {Rafieferantsoa} M.~H.,  {Thompson} R.~J.,   {Hopkins} P.~F.,
  2017, \mn@doi [\mnras] {10.1093/mnras/stx108}, \href
  {http://adsabs.harvard.edu/abs/2017MNRAS.467..115D} {467, 115}

\bibitem[\protect\citeauthoryear{{Di Cintio}, {Brook}, {Macci{\`o}}, {Stinson},
  {Knebe}, {Dutton}  \& {Wadsley}}{{Di Cintio} et~al.}{2014}]{dicintio14}
{Di Cintio} A.,  {Brook} C.~B.,  {Macci{\`o}} A.~V.,  {Stinson} G.~S.,  {Knebe}
  A.,  {Dutton} A.~A.,   {Wadsley} J.,  2014, \mn@doi [\mnras]
  {10.1093/mnras/stt1891}, \href
  {http://adsabs.harvard.edu/abs/2014MNRAS.437..415D} {437, 415}

\bibitem[\protect\citeauthoryear{{Dubois}, {Gavazzi}, {Peirani}  \&
  {Silk}}{{Dubois} et~al.}{2013}]{dubois13}
{Dubois} Y.,  {Gavazzi} R.,  {Peirani} S.,   {Silk} J.,  2013, \mn@doi [\mnras]
  {10.1093/mnras/stt997}, \href
  {http://adsabs.harvard.edu/abs/2013MNRAS.433.3297D} {433, 3297}

\bibitem[\protect\citeauthoryear{{Dubois} et~al.,}{{Dubois}
  et~al.}{2014}]{dubois14}
{Dubois} Y.,  et~al., 2014, \mn@doi [\mnras] {10.1093/mnras/stu1227}, \href
  {http://adsabs.harvard.edu/abs/2014MNRAS.444.1453D} {444, 1453}

\bibitem[\protect\citeauthoryear{{Dubois}, {Volonteri}, {Silk}, {Devriendt},
  {Slyz}  \& {Teyssier}}{{Dubois} et~al.}{2015}]{dubois15}
{Dubois} Y.,  {Volonteri} M.,  {Silk} J.,  {Devriendt} J.,  {Slyz} A.,
  {Teyssier} R.,  2015, \mn@doi [\mnras] {10.1093/mnras/stv1416}, \href
  {http://adsabs.harvard.edu/abs/2015MNRAS.452.1502D} {452, 1502}

\bibitem[\protect\citeauthoryear{{Dubois}, {Peirani}, {Pichon}, {Devriendt},
  {Gavazzi}, {Welker}  \& {Volonteri}}{{Dubois} et~al.}{2016}]{dubois16}
{Dubois} Y.,  {Peirani} S.,  {Pichon} C.,  {Devriendt} J.,  {Gavazzi} R.,
  {Welker} C.,   {Volonteri} M.,  2016, \mn@doi [\mnras]
  {10.1093/mnras/stw2265}, \href
  {http://adsabs.harvard.edu/abs/2016MNRAS.463.3948D} {463, 3948}

\bibitem[\protect\citeauthoryear{{El-Badry}, {Wetzel}, {Geha}, {Hopkins},
  {Kere{\v s}}, {Chan}  \& {Faucher-Gigu{\`e}re}}{{El-Badry}
  et~al.}{2016}]{elbadry16}
{El-Badry} K.,  {Wetzel} A.,  {Geha} M.,  {Hopkins} P.~F.,  {Kere{\v s}} D.,
  {Chan} T.~K.,   {Faucher-Gigu{\`e}re} C.-A.,  2016, \mn@doi [\apj]
  {10.3847/0004-637X/820/2/131}, \href
  {http://adsabs.harvard.edu/abs/2016ApJ...820..131E} {820, 131}

\bibitem[\protect\citeauthoryear{{Federrath}}{{Federrath}}{2015}]{federrath15}
{Federrath} C.,  2015, \mn@doi [\mnras] {10.1093/mnras/stv941}, \href
  {http://adsabs.harvard.edu/abs/2015MNRAS.450.4035F} {450, 4035}

\bibitem[\protect\citeauthoryear{{Ferland} et~al.,}{{Ferland}
  et~al.}{2013}]{ferland13}
{Ferland} G.~J.,  et~al., 2013, \rmxaa, \href
  {http://adsabs.harvard.edu/abs/2013RMxAA..49..137F} {49, 137}

\bibitem[\protect\citeauthoryear{{Ferreras}, {B{\"o}hm}, {Ziegler}  \&
  {Silk}}{{Ferreras} et~al.}{2014}]{ferreras14}
{Ferreras} I.,  {B{\"o}hm} A.,  {Ziegler} B.,   {Silk} J.,  2014, \mn@doi
  [\mnras] {10.1093/mnras/stt2018}, \href
  {http://adsabs.harvard.edu/abs/2014MNRAS.437.1872F} {437, 1872}

\bibitem[\protect\citeauthoryear{{Gallazzi}, {Charlot}, {Brinchmann}, {White}
  \& {Tremonti}}{{Gallazzi} et~al.}{2005}]{gallazzi05}
{Gallazzi} A.,  {Charlot} S.,  {Brinchmann} J.,  {White} S.~D.~M.,   {Tremonti}
  C.~A.,  2005, \mn@doi [\mnras] {10.1111/j.1365-2966.2005.09321.x}, \href
  {http://adsabs.harvard.edu/abs/2005MNRAS.362...41G} {362, 41}

\bibitem[\protect\citeauthoryear{{Gatto} et~al.,}{{Gatto}
  et~al.}{2017}]{gatto17}
{Gatto} A.,  et~al., 2017, \mn@doi [\mnras] {10.1093/mnras/stw3209}, \href
  {http://adsabs.harvard.edu/abs/2017MNRAS.466.1903G} {466, 1903}

\bibitem[\protect\citeauthoryear{{Geen}, {Hennebelle}, {Tremblin}  \&
  {Rosdahl}}{{Geen} et~al.}{2016}]{geen16}
{Geen} S.,  {Hennebelle} P.,  {Tremblin} P.,   {Rosdahl} J.,  2016, \mn@doi
  [\mnras] {10.1093/mnras/stw2235}, \href
  {http://adsabs.harvard.edu/abs/2016MNRAS.463.3129G} {463, 3129}

\bibitem[\protect\citeauthoryear{{Gingold} \& {Monaghan}}{{Gingold} \&
  {Monaghan}}{1982}]{gingold82}
{Gingold} R.~A.,  {Monaghan} J.~J.,  1982, \mn@doi [Journal of Computational
  Physics] {10.1016/0021-9991(82)90025-0}, \href
  {http://adsabs.harvard.edu/abs/1982JCoPh..46..429G} {46, 429}

\bibitem[\protect\citeauthoryear{{Gonz{\'a}lez Delgado} et~al.,}{{Gonz{\'a}lez
  Delgado} et~al.}{2014}]{delgado14}
{Gonz{\'a}lez Delgado} R.~M.,  et~al., 2014, \mn@doi [\apjl]
  {10.1088/2041-8205/791/1/L16}, \href
  {http://adsabs.harvard.edu/abs/2014ApJ...791L..16G} {791, L16}

\bibitem[\protect\citeauthoryear{{Governato} et~al.,}{{Governato}
  et~al.}{2010}]{governato10}
{Governato} F.,  et~al., 2010, \mn@doi [\nat] {10.1038/nature08640}, \href
  {http://adsabs.harvard.edu/abs/2010Natur.463..203G} {463, 203}

\bibitem[\protect\citeauthoryear{{Haardt} \& {Madau}}{{Haardt} \&
  {Madau}}{2012}]{haardt12}
{Haardt} F.,  {Madau} P.,  2012, \mn@doi [\apj] {10.1088/0004-637X/746/2/125},
  \href {http://adsabs.harvard.edu/abs/2012ApJ...746..125H} {746, 125}

\bibitem[\protect\citeauthoryear{{Hahn} \& {Abel}}{{Hahn} \&
  {Abel}}{2013}]{hahn13}
{Hahn} O.,  {Abel} T.,  2013, {MUSIC: MUlti-Scale Initial Conditions},
  Astrophysics Source Code Library (\mn@eprint {ascl} {1311.011})

\bibitem[\protect\citeauthoryear{{Harris}, {Harris}  \& {Alessi}}{{Harris}
  et~al.}{2013}]{harris13}
{Harris} W.~E.,  {Harris} G.~L.~H.,   {Alessi} M.,  2013, \mn@doi [\apj]
  {10.1088/0004-637X/772/2/82}, \href
  {http://adsabs.harvard.edu/abs/2013ApJ...772...82H} {772, 82}

\bibitem[\protect\citeauthoryear{{Hopkins}}{{Hopkins}}{2015}]{hopkins15}
{Hopkins} P.~F.,  2015, \mn@doi [\mnras] {10.1093/mnras/stv195}, \href
  {http://adsabs.harvard.edu/abs/2015MNRAS.450...53H} {450, 53}

\bibitem[\protect\citeauthoryear{{Hopkins}, {Kere{\v s}}, {O{\~n}orbe},
  {Faucher-Gigu{\`e}re}, {Quataert}, {Murray}  \& {Bullock}}{{Hopkins}
  et~al.}{2014}]{hopkins14}
{Hopkins} P.~F.,  {Kere{\v s}} D.,  {O{\~n}orbe} J.,  {Faucher-Gigu{\`e}re}
  C.-A.,  {Quataert} E.,  {Murray} N.,   {Bullock} J.~S.,  2014, \mn@doi
  [\mnras] {10.1093/mnras/stu1738}, \href
  {http://adsabs.harvard.edu/abs/2014MNRAS.445..581H} {445, 581}

\bibitem[\protect\citeauthoryear{{Hopkins} et~al.,}{{Hopkins}
  et~al.}{2017}]{hopkins17}
{Hopkins} P.~F.,  et~al., 2017, preprint, \href
  {http://adsabs.harvard.edu/abs/2017arXiv170206148H} {} (\mn@eprint {arXiv}
  {1702.06148})

\bibitem[\protect\citeauthoryear{{Hurley}, {Pols}  \& {Tout}}{{Hurley}
  et~al.}{2000}]{hurley00}
{Hurley} J.~R.,  {Pols} O.~R.,   {Tout} C.~A.,  2000, \mn@doi [\mnras]
  {10.1046/j.1365-8711.2000.03426.x}, \href
  {http://adsabs.harvard.edu/abs/2000MNRAS.315..543H} {315, 543}

\bibitem[\protect\citeauthoryear{{Ichikawa}, {Kajisawa}  \&
  {Akhlaghi}}{{Ichikawa} et~al.}{2012}]{ichikawa12}
{Ichikawa} T.,  {Kajisawa} M.,   {Akhlaghi} M.,  2012, \mn@doi [\mnras]
  {10.1111/j.1365-2966.2012.20674.x}, \href
  {http://adsabs.harvard.edu/abs/2012MNRAS.422.1014I} {422, 1014}

\bibitem[\protect\citeauthoryear{{Kalirai}, {Hansen}, {Kelson}, {Reitzel},
  {Rich}  \& {Richer}}{{Kalirai} et~al.}{2008}]{kalirai08}
{Kalirai} J.~S.,  {Hansen} B.~M.~S.,  {Kelson} D.~D.,  {Reitzel} D.~B.,  {Rich}
  R.~M.,   {Richer} H.~B.,  2008, \mn@doi [\apj] {10.1086/527028}, \href
  {http://adsabs.harvard.edu/abs/2008ApJ...676..594K} {676, 594}

\bibitem[\protect\citeauthoryear{{Katz}}{{Katz}}{1989}]{katz89}
{Katz} N.~S.,  1989, PhD thesis, Princeton Univ., NJ.

\bibitem[\protect\citeauthoryear{{Keller}, {Wadsley}, {Benincasa}  \&
  {Couchman}}{{Keller} et~al.}{2014}]{keller14}
{Keller} B.~W.,  {Wadsley} J.,  {Benincasa} S.~M.,   {Couchman} H.~M.~P.,
  2014, \mn@doi [\mnras] {10.1093/mnras/stu1058}, \href
  {http://adsabs.harvard.edu/abs/2014MNRAS.442.3013K} {442, 3013}

\bibitem[\protect\citeauthoryear{{Kennicutt}}{{Kennicutt}}{1998}]{kennicutt98}
{Kennicutt} Jr. R.~C.,  1998, \mn@doi [\apj] {10.1086/305588}, \href
  {http://adsabs.harvard.edu/abs/1998ApJ...498..541K} {498, 541}

\bibitem[\protect\citeauthoryear{{Kim} \& {Ostriker}}{{Kim} \&
  {Ostriker}}{2015}]{kimc15}
{Kim} C.-G.,  {Ostriker} E.~C.,  2015, \mn@doi [\apj]
  {10.1088/0004-637X/802/2/99}, \href
  {http://adsabs.harvard.edu/abs/2015ApJ...802...99K} {802, 99}

\bibitem[\protect\citeauthoryear{{Kim} et~al.,}{{Kim} et~al.}{2014}]{kim14}
{Kim} J.-h.,  et~al., 2014, \mn@doi [\apjs] {10.1088/0067-0049/210/1/14}, \href
  {http://adsabs.harvard.edu/abs/2014ApJS..210...14K} {210, 14}

\bibitem[\protect\citeauthoryear{{Kirby}, {Cohen}, {Guhathakurta}, {Cheng},
  {Bullock}  \& {Gallazzi}}{{Kirby} et~al.}{2013}]{kirby13}
{Kirby} E.~N.,  {Cohen} J.~G.,  {Guhathakurta} P.,  {Cheng} L.,  {Bullock}
  J.~S.,   {Gallazzi} A.,  2013, \mn@doi [\apj] {10.1088/0004-637X/779/2/102},
  \href {http://adsabs.harvard.edu/abs/2013ApJ...779..102K} {779, 102}

\bibitem[\protect\citeauthoryear{Knobel, Lilly, Woo  \& Kova{\v c}}{Knobel
  et~al.}{2015}]{knobel15}
Knobel C.,  Lilly S.~J.,  Woo J.,   Kova{\v c} K.,  2015, The Astrophysical
  Journal, 800, 24

\bibitem[\protect\citeauthoryear{{Knollmann} \& {Knebe}}{{Knollmann} \&
  {Knebe}}{2009}]{knollmann09}
{Knollmann} S.~R.,  {Knebe} A.,  2009, \mn@doi [\apjs]
  {10.1088/0067-0049/182/2/608}, \href
  {http://adsabs.harvard.edu/abs/2009ApJS..182..608K} {182, 608}

\bibitem[\protect\citeauthoryear{{Lehnert}, {van Driel}, {Le Tiran}, {Di
  Matteo}  \& {Haywood}}{{Lehnert} et~al.}{2015}]{lehnert15}
{Lehnert} M.~D.,  {van Driel} W.,  {Le Tiran} L.,  {Di Matteo} P.,   {Haywood}
  M.,  2015, \mn@doi [\aap] {10.1051/0004-6361/201322630}, \href
  {http://adsabs.harvard.edu/abs/2015A%26A...577A.112L} {577, A112}

\bibitem[\protect\citeauthoryear{{Leitner} \& {Kravtsov}}{{Leitner} \&
  {Kravtsov}}{2011}]{leitner11}
{Leitner} S.~N.,  {Kravtsov} A.~V.,  2011, \mn@doi [\apj]
  {10.1088/0004-637X/734/1/48}, \href
  {http://adsabs.harvard.edu/abs/2011ApJ...734...48L} {734, 48}

\bibitem[\protect\citeauthoryear{{Maiolino} et~al.,}{{Maiolino}
  et~al.}{2008}]{maiolino08}
{Maiolino} R.,  et~al., 2008, \mn@doi [\aap] {10.1051/0004-6361:200809678},
  \href {http://adsabs.harvard.edu/abs/2008A%26A...488..463M} {488, 463}

\bibitem[\protect\citeauthoryear{{Mannucci} et~al.,}{{Mannucci}
  et~al.}{2009}]{mannucci09}
{Mannucci} F.,  et~al., 2009, \mn@doi [\mnras]
  {10.1111/j.1365-2966.2009.15185.x}, \href
  {http://adsabs.harvard.edu/abs/2009MNRAS.398.1915M} {398, 1915}

\bibitem[\protect\citeauthoryear{{Maoz}, {Mannucci}  \& {Brandt}}{{Maoz}
  et~al.}{2012}]{maoz12}
{Maoz} D.,  {Mannucci} F.,   {Brandt} T.~D.,  2012, \mn@doi [\mnras]
  {10.1111/j.1365-2966.2012.21871.x}, \href
  {http://adsabs.harvard.edu/abs/2012MNRAS.426.3282M} {426, 3282}

\bibitem[\protect\citeauthoryear{{Mayer}, {Governato}  \& {Kaufmann}}{{Mayer}
  et~al.}{2008}]{mayer08}
{Mayer} L.,  {Governato} F.,   {Kaufmann} T.,  2008, Advanced Science Letters,
  \href {http://adsabs.harvard.edu/abs/2008ASL.....1....7M} {1, 7}

\bibitem[\protect\citeauthoryear{{McConnachie}}{{McConnachie}}{2012}]{mcconnachie12}
{McConnachie} A.~W.,  2012, \mn@doi [\aj] {10.1088/0004-6256/144/1/4}, \href
  {http://adsabs.harvard.edu/abs/2012AJ....144....4M} {144, 4}

\bibitem[\protect\citeauthoryear{{Oppenheimer}, {Dav{\'e}}, {Kere{\v s}},
  {Fardal}, {Katz}, {Kollmeier}  \& {Weinberg}}{{Oppenheimer}
  et~al.}{2010}]{oppenheimer10}
{Oppenheimer} B.~D.,  {Dav{\'e}} R.,  {Kere{\v s}} D.,  {Fardal} M.,  {Katz}
  N.,  {Kollmeier} J.~A.,   {Weinberg} D.~H.,  2010, \mn@doi [\mnras]
  {10.1111/j.1365-2966.2010.16872.x}, \href
  {http://adsabs.harvard.edu/abs/2010MNRAS.406.2325O} {406, 2325}

\bibitem[\protect\citeauthoryear{{Pakmor} \& {Springel}}{{Pakmor} \&
  {Springel}}{2013}]{pakmor13}
{Pakmor} R.,  {Springel} V.,  2013, \mn@doi [\mnras] {10.1093/mnras/stt428},
  \href {http://adsabs.harvard.edu/abs/2013MNRAS.432..176P} {432, 176}

\bibitem[\protect\citeauthoryear{{Pakmor}, {Pfrommer}, {Simpson}  \&
  {Springel}}{{Pakmor} et~al.}{2016}]{pakmor16}
{Pakmor} R.,  {Pfrommer} C.,  {Simpson} C.~M.,   {Springel} V.,  2016, \mn@doi
  [\apjl] {10.3847/2041-8205/824/2/L30}, \href
  {http://adsabs.harvard.edu/abs/2016ApJ...824L..30P} {824, L30}

\bibitem[\protect\citeauthoryear{{Peirani} et~al.,}{{Peirani}
  et~al.}{2016}]{peirani16}
{Peirani} S.,  et~al., 2016, preprint, \href
  {http://adsabs.harvard.edu/abs/2016arXiv161109922P} {} (\mn@eprint {arXiv}
  {1611.09922})

\bibitem[\protect\citeauthoryear{{Planck Collaboration} XIII}{{Planck
  Collaboration} XIII}{2016}]{planck16}
{Planck Collaboration} XIII, 2016, \mn@doi [A&A]
  {10.1051/0004-6361/201525830}, 594, A13

\bibitem[\protect\citeauthoryear{{Pontzen} \& {Governato}}{{Pontzen} \&
  {Governato}}{2012}]{pontzen12}
{Pontzen} A.,  {Governato} F.,  2012, \mn@doi [\mnras]
  {10.1111/j.1365-2966.2012.20571.x}, \href
  {http://adsabs.harvard.edu/abs/2012MNRAS.421.3464P} {421, 3464}

\bibitem[\protect\citeauthoryear{{Price}}{{Price}}{2007}]{splash07}
{Price} D.~J.,  2007, \mn@doi [\pasa] {10.1071/AS07022}, \href
  {http://adsabs.harvard.edu/abs/2007PASA...24..159P} {24, 159}

\bibitem[\protect\citeauthoryear{{Rosdahl}, {Schaye}, {Dubois}, {Kimm}  \&
  {Teyssier}}{{Rosdahl} et~al.}{2017}]{rosdahl17}
{Rosdahl} J.,  {Schaye} J.,  {Dubois} Y.,  {Kimm} T.,   {Teyssier} R.,  2017,
  \mn@doi [\mnras] {10.1093/mnras/stw3034}, \href
  {http://adsabs.harvard.edu/abs/2017MNRAS.466...11R} {466, 11}

\bibitem[\protect\citeauthoryear{{Ro{\v s}kar}, {Teyssier}, {Agertz},
  {Wetzstein}  \& {Moore}}{{Ro{\v s}kar} et~al.}{2014}]{roskar14}
{Ro{\v s}kar} R.,  {Teyssier} R.,  {Agertz} O.,  {Wetzstein} M.,   {Moore} B.,
  2014, \mn@doi [\mnras] {10.1093/mnras/stu1548}, \href
  {http://adsabs.harvard.edu/abs/2014MNRAS.444.2837R} {444, 2837}

\bibitem[\protect\citeauthoryear{{Salem}, {Bryan}  \& {Corlies}}{{Salem}
  et~al.}{2016}]{salem16}
{Salem} M.,  {Bryan} G.~L.,   {Corlies} L.,  2016, \mn@doi [\mnras]
  {10.1093/mnras/stv2641}, \href
  {http://adsabs.harvard.edu/abs/2016MNRAS.456..582S} {456, 582}

\bibitem[\protect\citeauthoryear{{Sales} et~al.,}{{Sales}
  et~al.}{2017}]{sales17}
{Sales} L.~V.,  et~al., 2017, \mn@doi [\mnras] {10.1093/mnras/stw2461}, \href
  {http://adsabs.harvard.edu/abs/2017MNRAS.464.2419S} {464, 2419}

\bibitem[\protect\citeauthoryear{{Schaye} et~al.,}{{Schaye}
  et~al.}{2015}]{schaye15}
{Schaye} J.,  et~al., 2015, \mn@doi [\mnras] {10.1093/mnras/stu2058}, \href
  {http://adsabs.harvard.edu/abs/2015MNRAS.446..521S} {446, 521}

\bibitem[\protect\citeauthoryear{{Simpson}, {Pakmor}, {Marinacci}, {Pfrommer},
  {Springel}, {Glover}, {Clark}  \& {Smith}}{{Simpson}
  et~al.}{2016}]{simpson16}
{Simpson} C.~M.,  {Pakmor} R.,  {Marinacci} F.,  {Pfrommer} C.,  {Springel} V.,
   {Glover} S.~C.~O.,  {Clark} P.~C.,   {Smith} R.~J.,  2016, \mn@doi [\apjl]
  {10.3847/2041-8205/827/2/L29}, \href
  {http://adsabs.harvard.edu/abs/2016ApJ...827L..29S} {827, L29}

\bibitem[\protect\citeauthoryear{{Somerville}, {Popping}  \&
  {Trager}}{{Somerville} et~al.}{2015}]{somerville15}
{Somerville} R.~S.,  {Popping} G.,   {Trager} S.~C.,  2015, \mn@doi [\mnras]
  {10.1093/mnras/stv1877}, \href
  {http://adsabs.harvard.edu/abs/2015MNRAS.453.4337S} {453, 4337}

\bibitem[\protect\citeauthoryear{{Springel}}{{Springel}}{2005}]{springel05}
{Springel} V.,  2005, \mn@doi [\mnras] {10.1111/j.1365-2966.2005.09655.x},
  \href {http://adsabs.harvard.edu/abs/2005MNRAS.364.1105S} {364, 1105}

\bibitem[\protect\citeauthoryear{{Springel}}{{Springel}}{2010}]{springel10}
{Springel} V.,  2010, \mn@doi [\mnras] {10.1111/j.1365-2966.2009.15715.x},
  \href {http://adsabs.harvard.edu/abs/2010MNRAS.401..791S} {401, 791}

\bibitem[\protect\citeauthoryear{{Stinson}, {Seth}, {Katz}, {Wadsley},
  {Governato}  \& {Quinn}}{{Stinson} et~al.}{2006}]{stinson06}
{Stinson} G.,  {Seth} A.,  {Katz} N.,  {Wadsley} J.,  {Governato} F.,   {Quinn}
  T.,  2006, \mn@doi [\mnras] {10.1111/j.1365-2966.2006.11097.x}, \href
  {http://adsabs.harvard.edu/abs/2006MNRAS.373.1074S} {373, 1074}

\bibitem[\protect\citeauthoryear{{Teyssier}}{{Teyssier}}{2002}]{teyssier02}
{Teyssier} R.,  2002, \mn@doi [\aap] {10.1051/0004-6361:20011817}, \href
  {http://adsabs.harvard.edu/abs/2002A%26A...385..337T} {385, 337}

\bibitem[\protect\citeauthoryear{{Teyssier}, {Pontzen}, {Dubois}  \&
  {Read}}{{Teyssier} et~al.}{2013}]{teyssier13}
{Teyssier} R.,  {Pontzen} A.,  {Dubois} Y.,   {Read} J.~I.,  2013, \mn@doi
  [\mnras] {10.1093/mnras/sts563}, \href
  {http://adsabs.harvard.edu/abs/2013MNRAS.429.3068T} {429, 3068}

\bibitem[\protect\citeauthoryear{{Tollet} et~al.,}{{Tollet}
  et~al.}{2016}]{tollet16}
{Tollet} E.,  et~al., 2016, \mn@doi [\mnras] {10.1093/mnras/stv2856}, \href
  {http://adsabs.harvard.edu/abs/2016MNRAS.456.3542T} {456, 3542}

\bibitem[\protect\citeauthoryear{{Torrey}, {Vogelsberger}, {Sijacki},
  {Springel}  \& {Hernquist}}{{Torrey} et~al.}{2012}]{torrey12}
{Torrey} P.,  {Vogelsberger} M.,  {Sijacki} D.,  {Springel} V.,   {Hernquist}
  L.,  2012, \mn@doi [\mnras] {10.1111/j.1365-2966.2012.22082.x}, \href
  {http://adsabs.harvard.edu/abs/2012MNRAS.427.2224T} {427, 2224}

\bibitem[\protect\citeauthoryear{{Vogelsberger} et~al.,}{{Vogelsberger}
  et~al.}{2014}]{vogelsberger14}
{Vogelsberger} M.,  et~al., 2014, \mn@doi [\mnras] {10.1093/mnras/stu1536},
  \href {http://adsabs.harvard.edu/abs/2014MNRAS.444.1518V} {444, 1518}

\bibitem[\protect\citeauthoryear{{Voit} \& {Donahue}}{{Voit} \&
  {Donahue}}{2011}]{voit11}
{Voit} G.~M.,  {Donahue} M.,  2011, \mn@doi [\apjl]
  {10.1088/2041-8205/738/2/L24}, \href
  {http://adsabs.harvard.edu/abs/2011ApJ...738L..24V} {738, L24}

\bibitem[\protect\citeauthoryear{{Volonteri}, {Dubois}, {Pichon}  \&
  {Devriendt}}{{Volonteri} et~al.}{2016}]{volonteri16}
{Volonteri} M.,  {Dubois} Y.,  {Pichon} C.,   {Devriendt} J.,  2016, \mn@doi
  [\mnras] {10.1093/mnras/stw1123}, \href
  {http://adsabs.harvard.edu/abs/2016MNRAS.460.2979V} {460, 2979}

\bibitem[\protect\citeauthoryear{{Woosley} \& {Heger}}{{Woosley} \&
  {Heger}}{2007}]{woosley07}
{Woosley} S.~E.,  {Heger} A.,  2007, \mn@doi [\physrep]
  {10.1016/j.physrep.2007.02.009}, \href
  {http://adsabs.harvard.edu/abs/2007PhR...442..269W} {442, 269}

\bibitem[\protect\citeauthoryear{{Yabe} et~al.,}{{Yabe} et~al.}{2014}]{yabe14}
{Yabe} K.,  et~al., 2014, \mn@doi [\mnras] {10.1093/mnras/stt2185}, \href
  {http://adsabs.harvard.edu/abs/2014MNRAS.437.3647Y} {437, 3647}

\bibitem[\protect\citeauthoryear{{Zibetti}, {Charlot}  \& {Rix}}{{Zibetti}
  et~al.}{2009}]{zibetti09}
{Zibetti} S.,  {Charlot} S.,   {Rix} H.-W.,  2009, \mn@doi [\mnras]
  {10.1111/j.1365-2966.2009.15528.x}, \href
  {http://adsabs.harvard.edu/abs/2009MNRAS.400.1181Z} {400, 1181}

\bibitem[\protect\citeauthoryear{{van der Wel} et~al.,}{{van der Wel}
  et~al.}{2014}]{vanderwel14}
{van der Wel} A.,  et~al., 2014, \mn@doi [\apj] {10.1088/0004-637X/788/1/28},
  \href {http://adsabs.harvard.edu/abs/2014ApJ...788...28V} {788, 28}

\makeatother
\end{thebibliography}
\label{lastpage}
\end{document}